\documentclass[fleqn,usenatbib]{mnras}

\DeclareRobustCommand{\VAN}[3]{#2}
\let\VANthebibliography\thebibliography
\def\thebibliography{\DeclareRobustCommand{\VAN}[3]{##3}\VANthebibliography}

\usepackage{graphicx}	%
\usepackage{amsmath}	%
\usepackage{amssymb}	%

\usepackage{newtxtext}

\usepackage{hyperref}
\usepackage{fixme}
\fxsetup{
  status=draft,
  author=CC,
  layout=inline,nomargin,nofootnote,
  theme=colorsig,
}
\FXRegisterAuthor{cc}{acc}{CC}
\FXRegisterAuthor{mrk}{amrk}{MRK}
\FXRegisterAuthor{bdw}{abdw}{BDW}

\newcommand{\newnew}[1]{{#1}}

\newcommand{\intg}{\int_{\nu_{g-}}^{\nu_{g+}}}

\newcommand{\nuplus}{\nu_{g+}}
\newcommand{\numinus}{\nu_{g-}}
\newcommand{\quokka}[0]{\textsc{quokka}}

\newcommand{\vecx}{\boldsymbol{x}}
\newcommand{\vecv}{\boldsymbol{v}}
\newcommand{\vecn}{\boldsymbol{n}}
\newcommand{\vecG}{\boldsymbol{G}}
\newcommand{\vecF}{\boldsymbol{F}}

\newcommand{\tenP}{\mathsf{P}}

\newcommand{\matU}{\mathbfss{U}}
\newcommand{\matF}{\mathbfss{F}}
\newcommand{\matS}{\mathbfss{S}}
\newcommand{\matT}{\mathbfss{T}}
\newcommand{\tenI}{\mathsf{I}}

\newcommand{\chat}{\hat{c}}

\newcommand{\aref}[1]{\hyperref[#1]{Appendix~\ref{#1}}}

\title[\textsc{quokka}: Multigroup RHD]{A novel numerical method for mixed-frame multigroup radiation-hydrodynamics with GPU acceleration implemented in the \textsc{quokka} code}

\author[He, Wibking, \& Krumholz]{
Chong-Chong He$^{1}$\thanks{E-mail: Chongchong.He@anu.edu.au (CCH)}, Benjamin D. Wibking$^{2}$, and Mark R. Krumholz$^{1,3}$
\\
$^{1}$Research School of Astronomy and Astrophysics, Australian National University, Canberra, ACT 2611, Australia\\
$^{2}$Department of Physics and Astronomy, Michigan State University, East Lansing, MI 48824, USA\\
$^{3}$ARC Centre of Excellence for Astronomy in Three Dimensions (ASTRO-3D), Canberra, ACT 2611, Australia
}

\date{Accepted XXX. Received YYY; in original form ZZZ}

\pubyear{2024}

\begin{document}
\label{firstpage}
\pagerange{\pageref{firstpage}--\pageref{lastpage}}
\maketitle

\begin{abstract}
Mixed-frame formulations of radiation-hydrodynamics (RHD), where the radiation quantities are computed in an inertial frame but matter quantities are in a comoving frame, are advantageous because they admit algorithms that conserve energy and momentum to machine precision and combine more naturally with adaptive mesh techniques, since unlike pure comoving-frame methods they do not face the problem that radiation quantities must change frame every time a cell is refined or coarsened. However, implementing multigroup RHD in a mixed-frame formulation presents challenges due to the complexity of handling frequency-dependent interactions and the Doppler shift of radiation boundaries. In this paper, we introduce a novel method for multigroup RHD that integrates a mixed-frame formulation with a piecewise powerlaw approximation for frequency dependence within groups. This approach ensures the exact conservation of total energy and momentum while effectively managing the Lorentz transformation of group boundaries and evaluation of group-averaged opacities. Our method takes advantage of the locality of matter-radiation coupling, allowing the source term for $N_g$ frequency groups to be handled with simple equations with a sparse Jacobian matrix of size $N_g + 1$, which can be inverted with $O(N_g)$ complexity. This results in a computational complexity that scales linearly with $N_g$ and requires no more communication than a pure hydrodynamics update, making it highly efficient for massively parallel and GPU-based systems. We implement our method in the GPU-accelerated RHD code \textsc{quokka} and demonstrate that it passes a wide range of numerical tests, \newnew{including preserving the asymptotic diffusion limit}. We demonstrate that the piecewise powerlaw method shows significant advantages over traditional opacity averaging methods for handling rapidly variable opacities with modest frequency resolution.
\end{abstract}


\section{Introduction}
\label{sec:intro}

\defcitealias{Wibking2022}{Paper~I}
\defcitealias{He2024}{Paper~II}

Radiation-hydrodynamics (RHD) plays a crucial role in modeling astrophysical phenomena where radiation is reponsible for transporting energy or momentum. Accurate treatment of radiation transport is essential for capturing these interactions accurately. While grey approximations that integrate over radiation frequency and only follow frequency-integrated quantities have been widely employed, they often fail to capture the complex  spectral features and frequency-dependent radiative processes that can significantly influence the system's evolution. For instance, only frequency-dependent methods can correctly model the absorption of ultraviolet radiation from stars by surrounding dust and its subsequent re-emission in the infrared, a phenomenon that is crucial to mediating how radiation pressure regulates star formation \citep[e.g.,][]{Rosen16a, Menon22c}. Moreover, multigroup formalisms have been shown to significantly impact the structure of radiative shocks compared to grey methods \citep{Vaytet2013} and to alter energy transport in stellar atmospheres \citep{Chiavassa2011}. 

In this context, multigroup RHD has emerged as a powerful tool, offering a more comprehensive and accurate representation of the radiation field by dividing the spectrum into multiple energy groups. Several works have applied multigroup RHD in comoving frame formulations  \citep{Vaytet2011, Zhang2013, Skinner2019}, meaning that radiation quantities are evaluated in the frame comoving with the matter. In the comoving frame, the matter-radiation interaction is simplified, with the Doppler effect manifesting as an advection of radiation quantities between groups related to the matter's velocity gradient.  However, the comoving frame formulation does not conserve total energy or momentum exactly, since the RHD equations in the comoving frame are not manifestly conservative and do not allow for the construction of conservative update schemes; instead, in such methods conservation is generally achieved only to order $v/c$. %

These conservation errors can be amplified in simulations using adaptive mesh refinement (AMR), where volumes of the simulation domain must be coarsened and refined. This creates a problem for comoving frame formulations of RHD: the children of a parent cell that is being refined generally have different velocities than the parent cell, which means that the radiation quantities associated with the child and parent cells are not defined in the same reference frame. One can ignore this difference -- this is the most common practice in comoving frame RHD-AMR codes -- but this incurs an error in conservation of order $v/c$ at each refinement stage, and in deeply nested calculations with many levels of refinement these errors may well add constructively, leading to an overall violation of conservation at unacceptable levels. In a single-group calculation one could in principle address this problem by explicitly transforming radiation quantities to the lab frame before coarsening and refinement operations, then transforming back afterwards, but this option is not available in a multigroup calculation: while the radiation energy and flux integrated over all frequencies form a four-vector that can be straightforwardly Lorentz-transformed, the radiation energy and flux integrated over a finite range of frequencies do not. Instead, the transformation of the flux between frames depends on the third moment of the specific intensity, requiring the adoption of an additional closure relation. Indeed, one is required to estimate the third moment of the intensity in \textit{any} comoving frame treatment of RHD  \citep{Mihalas1984, Lowrie1999, Hubeny2007}. In addition to adding an extra physical assumption to the problem, this step comes at potentially substantial computational cost \citep[e.g., see the discussion of the steps required to extend the $M_1$ closure to the third moment in][]{Vaytet2011}.

We recently released the \textsc{quokka}\footnote{Quadrilateral, Umbra-producing, Orthogonal, Kangaroo-conserving Kode
for Astrophysics! \href{https://github.com/quokka-astro/quokka}{https://github.com/quokka-astro/quokka}} code, a new GPU-accelerated AMR RHD code (\citealt{Wibking2022}, hereafter \citetalias{Wibking2022}) featuring a novel asymptotic-preserving time integration scheme (\citealt*{He2024}, hereafter \citetalias{He2024}) that conserves energy and momentum to machine precision and correctly recovers all limits of RHD. In \textsc{quokka} we solve the moment equations for RHD in the mixed-frame formulation, where the radiative quantities are defined in the lab frame (i.e.~Eulerian simulation coordinates), and the emissivity and absorption are described in the comoving frame of the fluid, where they can be assumed to be isotropic. This approach takes advantage of the simplicity of the hyperbolic operators in the lab frame, allowing for the exact conservation of total energy and momentum, while benefiting from the isotropic nature of matter emissivities and opacities in the comoving frame \citep[see][]{Castor2009}. It also integrates natively with AMR, since the lab frame radiation energy and flux are conserved quantities that can be coarsened or refined exactly like the conserved hydrodynamic quantities. In this formulation relativistic corrections to the opacity due to Doppler effects between the laboratory frame (where the radiation variables are defined) and the comoving frame (where microscopic interactions occur) appear as additional terms of first or higher order in $v/c$ in the rate of matter-radiation exchange \citep{Mihalas1984}. These terms must be retained for accuracy, but are relatively straightforward to handle numerically.

However, extending mixed-frame formulations to multigroup RHD remains a frontier problem. While at first it might seem simple to extend the mixed-frame approach by evolving the radiation moments for each frequency group in the lab frame while calculating group opacities/emissivities in the comoving frame, thereby retaining the advantages of the hyperbolic nature of the radiation transport in the lab frame while accounting for the frequency dependence of opacities in the comoving frame, a difficulty arises with the frequency boundaries. In a mixed-frame formulation these are defined in the lab frame, but they are Doppler shifted in the comoving frame where matter emissivities and opacities are defined. This necessitates careful calculation of group-mean opacities since they are only well-defined in the comoving frame. As a result of this difficulty, the only lab-frame multigroup RHD method currently in use in astrophysics is based on direct discretisation of the radiative transfer equation rather than solution of the moment equations \citep{Jiang22a}, an approach that is considerably more expensive in terms of both computation and memory. Moreover, while in principle this could be combined with the reduced speed of light approximation to yield an explicit method that would run efficiently on parallel GPUs, the only available current implementation in astrophysics relies on a global implicit update that does not.

In this paper, we tackle this problem and extend the \textsc{quokka} algorithm to include frequency dependence via the multigroup approach. The extended algorithm maintains all the desirable properties of the original scheme -- machine-precision conservation both during updates and when adaptively refining and coarsening, correct recovery of all asymptotic RHD limits -- while also allowing frequency-dependent opacities with a flexible decomposition of the frequency space in a conservative manner. This paper is organized as follows. In \autoref{sec:mgrhd} we examine the multigroup RHD equations in the mixed frame formulation, and develop our strategy for frequency discretisation. In \autoref{sec:numerical} we describe our numerical approach to implementing the scheme. \autoref{sec:tests} presents tests of the scheme. Finally, we summarise and discuss future prospects in \autoref{sec:conclusion}.

\section{Formulation of the multigroup RHD system}
\label{sec:mgrhd}

\begin{table}
    \centering
    \caption{Symbols used in this paper.}
    \begin{tabular}{p{0.15\columnwidth}p{0.4\columnwidth}p{0.28\columnwidth}}
    \hline
         Symbol & Meaning  &Units\\
         \hline
 $\rho$&gas density &${\rm g\;cm^{-3}}$\\
 $\vecv$, $v^i$&gas velocity &${\rm cm\;s^{-1}}$\\
 $p$ &gas pressure &${\rm erg\;cm^{-3}}$\\
         $E_{\rm gas}$& 
    gas total energy &${\rm erg\;cm^{-3}}$\\
 $E_{\rm kin}$&gas kinetic energy &${\rm erg\;cm^{-3}}$\\
 $\nu$& lab-frame radiation frequency&Hz\\
 $\numinus$, $\nuplus$& the lower and upper boundaries of the $g$th radiation group in lab frame&Hz\\
 $r_g$ & $=\nuplus/\numinus$ & Dimensionless \\
 $\nu_0$& comoving-frame radiation frequency&Hz\\
 $E_\nu$&specific radiation energy &${\rm erg\;cm^{-3}\;Hz^{-1}}$\\
 $E_g$ &$=\intg E_{\nu} d\nu$ &${\rm erg\;cm^{-3}}$\\
 $\vecF_{\nu}$ or $F_{\nu}^i$&specific radiation flux &${\rm erg\;cm^{-2}\;s^{-1}\;Hz^{-1}}$\\
 $\vecF_{g}$ or $F_{g}^i$&$=\intg \vecF_{\nu} d\nu$ or $\intg F_{\nu}^i d\nu$&${\rm erg\;cm^{-2}\;s^{-1}}$\\
 $\tenP_{\nu}$ or $P_{\nu}^{ij}$&specific radiation pressure tensor &${\rm erg\;cm^{-3}\;Hz^{-1}}$\\
 $\tenP_{g}$, $P_{g}^{ij}$ &$=\intg \tenP_{\nu} d\nu$ or $\intg P_{\nu}^{ij} d\nu$&${\rm erg\;cm^{-3}}$\\
 $G_{\nu}^0$&time-like component of the specific radiation four-force &${\rm erg\;cm^{-4}\;Hz^{-1}}$\\
 $G_{g}^0$&$=\intg G_{\nu}^0 d\nu$&${\rm erg\;cm^{-4}}$\\
 $\vecG_{\nu}$ or $G_{\nu}^i$&space-like components of the specific radiation four-force &${\rm erg\;cm^{-4}\;Hz^{-1}}$\\
 $\vecG_{g}$ or $G_{g}^i$&$=\intg \vecG_{\nu} d\nu$ or $\intg G_{\nu}^i d\nu$&${\rm erg\;cm^{-4}}$\\
 $\vecn$, $n^i$ & the unit vector (or its $i$th component) in Cartesian coordinates& Dimensionless\\
 $N_g$& number of photon groups&\\
 $\chi_0$&Comoving-frame absorption coefficient&${\rm cm^{-1}}$\\
 $\chi$& Lab-frame absorption coefficient&${\rm cm^{-1}}$\\
 $\eta_0$& Comoving-frame emissivity&${\rm erg\;cm^{-3}\;s^{-1}\;Hz^{-1}}$\\
 $\eta$& Lab-frame emissivity&${\rm erg\;cm^{-3}\;s^{-1}\;Hz^{-1}}$\\
 $B_{\nu}$&Planck function &${\rm erg\;cm^{-2}\;s^{-1}\;Hz^{-1}}$\\
 $B_g$&=$\intg B_{\nu} d\nu$ &${\rm erg\;cm^{-2}\;s^{-1}}$\\
 $R_g$& the $g$th-group source term&${\rm erg\;cm^{-3}~Hz^{-1}}$\\
 $c$&speed of light &${\rm cm\;s^{-1}}$\\
 $\chat$ &reduced speed of light &${\rm cm\;s^{-1}}$\\
 $\Delta_g(Q)$&$=Q(\nuplus) - Q(\numinus)$&\\
 $\chi_{0Q,g}$& Group-averaged absorption coefficient (\autoref{eq:chiQg})&${\rm cm^{-1}}$\\
 $\alpha_{\chi_0,g}$& Powerlaw index of the absorption coefficient in group $g$& Dimensionless\\
 $\alpha_{Q,g}$& Powerlaw index of the radiation quantity $Q$ in group $g$& Dimensionless\\
 $a_R$& radiation constant&${\rm erg\;cm^{-3}~K^{-4}}$\\
 \hline
 \end{tabular}
 \label{tab:1}
\end{table}
In this section, we present the multigroup RHD equations that \quokka{} solves. We begin by deriving the multigroup RHD system of equations in \autoref{sec:RHD1}, and then introducing our strategy for evaluating group-averaged exchange terms in \autoref{sec:RHD2}. For the convenience of the readers, we list the notations and symbols used in this paper in \autoref{tab:1}. 

\subsection{The multigroup RHD equations}\label{sec:RHD1}

The full set of frequency-integrated RHD equations we solve has been introduced in \citetalias{He2024}. Here, we repeat these equations while writing the radiation-related terms in frequency-dependent form. These equations are 
\begin{align} \label{eq:hyper}
  \frac{\partial \matU}{\partial t}+\nabla \cdot\matF_{\matU}=\matS_{\matU},
\end{align}
where 
\begin{equation}
  \label{eq:G}  
  \matU=\left[
    \begin{array}{c}
      \rho \\
      \rho \boldsymbol{v} \\
      E_{\rm gas} \\
      E_{\nu} \\
      \frac{1}{c^2} \boldsymbol{F}_{\nu}
    \end{array}\right], \;
  \matF_{\matU} = \left[
    \begin{array}{c}
      \rho \boldsymbol{v} \\
      \rho \boldsymbol{v} \otimes \boldsymbol{v}+p \\
      (E_{\rm gas} + p) \boldsymbol{v} \\
      \boldsymbol{F}_{\nu} \\
      \tenP_{\nu}
    \end{array}\right], \;
  \matS_{\matU}=\left[
    \begin{array}{c}
      0 \\
      \int_0^{\infty} \boldsymbol{G}_{\nu} d\nu \\
      c \int_0^{\infty} G_{{\nu}}^0 d\nu \\
      - c G_{\nu}^0 \\
      - \boldsymbol{G}_{\nu}
    \end{array}\right]
\end{equation}
are vectors of the conserved quantities, the advection terms, and the source terms, respectively. In the equations above, $\rho$ is the matter density, $\vecv$ is the matter velocity, $p$ is the matter pressure, and $E_{\rm gas}$ is the gas total energy density;
$E_{\nu}$, $\vecF_{\nu}$, and $\tenP_{\nu}$ are, respectively, the specific radiation energy density, radiation flux, and radiation pressure tensor, which are defined in terms of the zeroth, first, and second moments of the specific intensity $I(\vecn, \nu)$ in direction $\vecn$ at frequency $\nu$ as
\begin{equation}
    \begin{split}
& E_{\nu} = c^{-1} \oint I(\vecn, \nu) d \Omega \\
& F_{\nu}^i = \oint n^i I( \vecn, \nu) d \Omega \\
& P_{\nu}^{ij} = c^{-1} \oint n^i n^j I(\mathbf{n}, \nu) d \Omega .
    \end{split}
\end{equation}
Note that all quantities in these expressions are in the lab frame. As discussed in the introduction, the lab-frame equations are manifestly conservative, a feature that makes it possible to build conservative algorithms relatively straightforwardly. 

Our next step is to write the frequency-dependent radiation four-force, $(G_{\nu}^0, \vecG_{\nu})$, in the mixed-frame formulation. We assume that the emitting matter is in local thermal equilibrium, so the source function in the comoving frame is the Planck function, and we neglect scattering. In the lab frame, we have \citep{Mihalas1984} 
\begin{equation}\label{eq:G0_1}
\begin{split}
    - c G^0_{\nu} \equiv \oint d \Omega ~[\eta(\vecn, \nu) - \chi(\vecn, \nu) I(\vecn, \nu)] \\
    - G^i_{\nu} \equiv c^{-1} \oint d \Omega ~n^i [\eta(\vecn, \nu) - \chi(\vecn, \nu) I(\vecn, \nu)],
\end{split}
\end{equation}
where $\eta(\vecn, \nu)$ and $\chi(\vecn, \nu)$ are the matter specific emissivity and absorption coefficient in the lab frame; note that, in the lab frame, these quantities depend on direction $\vecn$ due to relativistic boosting effects, which cannot be neglected even in sub-relativistic flows in a consistent theory of RHD. Following our mixed-frame approach, we now seek to rewrite the equations in terms of the comoving-frame emissivity $\eta_0(\nu)$ and absorption coefficient $\chi_0(\nu)$, which are simpler because they do not depend on direction.\footnote{Here and throughout we shall follow a convention whereby terms with a subscript 0 are evaluated in the comoving frame, while those without a subscript 0 are evaluated in the lab frame.} Integrated over all frequencies and for matter in local thermodynamic equilibrium, this transformation to mixed-frame is relatively simple; to first order $v/c$ it is (see \citealt{Mihalas2001}, and also Equation (3) and (4) of \citetalias{He2024}):

    \begin{align}
        -c G^0 &\equiv -\int_0^\infty c G^0_\nu \, d\nu \nonumber \\
        &= 4 \pi \chi_{0P} B - c \chi_{0E} E + c^{-1} (2 \chi_{0E} - \chi_{0F}) v^i F^i \label{eq:Gsingle1}\\
        -G^i &\equiv -\int_0^\infty G^i_\nu \, d\nu \nonumber \\
        & =  -c^{-1} \chi_{0F} F^i + 4 \pi c^{-2} v^i \chi_{0P} B        \nonumber \\
        & \quad {} + c^{-1} \chi_{0F} v^j P^{ji} + c^{-1} (\chi_{0F} - \chi_{0E}) v_i E, \label{eq:Gsingle2}
    \end{align}
where $B, E, F^i, P^{ij}$ are frequency-integrated Planck function, radiation energy density, radiation flux, and radiation pressure written in the lab frame, but now $\chi_{0P}, \chi_{0E}$ and $\chi_{0F}$ are the comoving-frame Planck-mean, energy-mean, and flux-mean opacity.\footnote{Note that we have assumed that the flux spectrum is the same in all directions, so that the direction-dependent $\chi_{0F}^{(i)}$ can be replaced by a scalar $\chi_{0F}$.} However, for the types of system where we might want to carry out a frequency-resolved calculation -- those far from thermodynamic equilibrium over at least part of the spatial or frequency domain -- these comoving-frame mean opacities are unknown, because they depend on the frequency distribution of the radiation quantities, which we know only in the lab frame. While in principle one could solve for the frequency distribution in the lab frame and then transform to the comoving frame to evaluate the opacities, this would remove the main advantage of the mixed-frame approach, which is avoiding such cumbersome transformations. 

We therefore proceed instead by following \cite{Mihalas1984}: to handle the lab-frame angle-frequency dependence of the absorption and emission terms, we expand the lab-frame frequency $\nu$ around the comoving-frame frequency $\nu_0$ to first order in $v/c$, as
\begin{equation}
    \nu / \nu_0 = 1 + \vecn \cdot \vecv / c.
\end{equation}
We can similarly expand the absorption coefficient and emissivity by writing down the Lorentz transformations for them and then expanding to $\mathrm{O}(v / c)$, yielding
\begin{equation}
\begin{split}
\chi(\vecn, \nu) & = (\nu_0/\nu) \chi_0(\nu_0) = \chi_0(\nu)-\left(\vecn \cdot \frac{\vecv}{c}\right)\left(\chi_0(\nu)+\nu\frac{\partial \chi_0}{\partial \nu}\right) \\
\eta(\vecn, \nu) &= (\nu/\nu_0)^2 \eta_0(\nu_0) = \eta_0(\nu)+\left(\vecn \cdot \frac{\vecv}{c}\right)\left(2 \eta_0(\nu)-\nu\frac{\partial \eta_0}{\partial \nu}\right) .
\end{split}
\end{equation}

Evaluating the integration over $d \Omega$ in \autoref{eq:G0_1} and applying our assumption that the matter is in LTE so $\eta_0(\nu) = \chi_0(\nu) B_{\nu}$, where $B_\nu$ is the Planck function evaluated at the local matter temperature, we obtain the time-like and space-like components of the radiation four-force (\citealt{Mihalas1978}, their Eqs. 2.19 and 2.21, see also \citealt{Mihalas1984}, their Eqs. 93.5 and 93.6)\footnote{We point out that, although these equations appear slightly different in form than the mixed-frame equations derived by \citet{Lowrie1999}, the apparent difference arises solely from the fact that we have expressed the four-force purely in terms of lab frame radiation quantities, while they express theirs in terms of comoving frame quantities. One can readily verify that our expressions are identical to order $v/c$ by substituting their equations 27a and 27b for the comoving frame energy and flux into their equations 28a and 28b for the time-like and space-like parts of the four-force, setting the scattering opacity to zero (since we neglect scattering), and dropping terms of order $v^2/c^2$ and higher.}
\begin{align} \label{eq:cG0}
    - c G_{\nu}^0 = 4 \pi \chi_0({\nu}) B_{\nu} - c \chi_{0}(\nu) E_{\nu} + \left(\chi_{0}(\nu) + \nu \frac{\partial \chi_{0}}{\partial \nu} \right) \frac{v^i F_{\nu}^i}{c},
\end{align}
\begin{equation}\label{eq:cGi}
\begin{split}
    - G_{\nu}^i &= - \chi_{0}(\nu) \frac{F_{\nu}^i}{c} + \frac{4}{3} \pi c^{-2} v^i \left( 2 \chi_0(\nu) B_{\nu} - \nu \frac{\partial (\chi_0(\nu) B_{\nu})}{\partial \nu} \right) \\
    & \quad {} + c^{-1} \left( \chi_0(\nu) + \nu \frac{\partial \chi_0}{\partial \nu} \right) v^j P_{\nu}^{ji}.
\end{split}
\end{equation}

We now divide the frequency domain into a finite ($N_g$) number of bins or groups. We define group-integrated radiation quantities as
\begin{equation}\label{eq:Qg}
    Q_{g} \equiv \intg Q_{\nu} d\nu,
\end{equation}
where $Q = B, E, \vecF, \tenP, G^{0}$, and $\vecG$, which represent the Planck function, the radiative energy, flux, and pressure, and the time-like and space-like components of the radiation four-force; $\numinus$ and $\nuplus$ are the frequency at the lower and upper boundaries, respectively, of the $g$th group, and groups are contiguous in frequency so $\nuplus = \nu_{g+1,-}$. If we now integrate the final two lines of \autoref{eq:G} over each of the groups, and assuming that the groups cover a broad enough frequency range that we can neglect contributions to the frequency-integrated four-force $(G^0, \vecG)$ from frequencies $<\nu_{1-}$ and $>\nu_{N_g+}$, the vectors of conserved quantities, advection terms, and source terms become
\begin{equation}
  \matU =\left[
    \begin{array}{c}
      \rho \\
      \rho \boldsymbol{v} \\
      E_{\rm gas} \\
      E_g \\
      \frac{1}{c^2} \boldsymbol{F}_g
    \end{array}\right], \;
  \matF_{\matU} = \left[
    \begin{array}{c}
      \rho \boldsymbol{v} \\
      \rho \boldsymbol{v} \otimes \boldsymbol{v}+p \\
      (E_{\rm gas} + p) \boldsymbol{v} \\
      \boldsymbol{F}_g \\
      \tenP_g
    \end{array}\right], \;
  \matS_{\matU}=\left[
    \begin{array}{c}
      0 \\
      \sum_g \boldsymbol{G}_g \\
      c \sum_g G^0_{g} \\
      - c G^0_{g} \\
      - \boldsymbol{G}_g
    \end{array}\right],
    \label{eq:MG_state_vec}
\end{equation}
where the group-integrated time-like component of the radiation four-force is
\begin{equation}\label{eq:G0_2new}
    \begin{split}
    - c G_{g}^0 &= \intg (- c G_{\nu}^0) d\nu \\
    &= 4 \pi \intg \chi_0({\nu}) B_{\nu} d\nu - c \intg \chi_{0}(\nu) E_{\nu} d\nu \\ 
    & \quad{} + c^{-1} v^i \intg \left[ \chi_0(\nu) + \nu \frac{\partial \chi_{0}}{\partial \nu} \right] F^i_{\nu} d \nu,
    \end{split}
\end{equation} 
and the group-integrated space-like components are 
\begin{equation}\label{eq:Gi_2new}
    \begin{split}
    - G_{g}^i &= \intg (- G_{\nu}^i) d\nu \\
    &= - c^{-1} \intg \chi_{0}(\nu) F^i_{\nu} d\nu \\
    & \quad{} + 4 \pi c^{-2} v^i \left[  \intg \chi_0(\nu) B_{\nu} d\nu -\frac{1}{3} \Delta_g (\nu \chi_0 B_{\nu}) \right] \\
    & \quad{} + c^{-1} v^j \intg \left[ \chi_0(\nu) + \nu \frac{\partial \chi_0}{\partial \nu}\right] P_{\nu}^{ji} d\nu.
    \end{split}
\end{equation}

The equations do require that one adopt a closure relation for the radiation pressure tensor $\tenP_{g}$. \quokka{} uses the \cite{Levermore84a} closure for the RHD system by default -- see \citetalias{Wibking2022} for full details -- but the method we describe here is independent of this choice. In our multigroup implementation we express the radiation pressure tensor for a radiation group, $\tenP_g$, in terms of the radiation energy density and flux from the same group, $E_g$ and $\vecF_g$, i.e., we apply this closure group-by-group.

Before proceeding further, we pause to demonstrate that our formulation of the radiation four-force to order $v/c$ is consistent. The reason we might worry that it is not is that, as \citet{Lowrie1999} point out, there can be an inconsistency in keeping only order $v/c$ terms in the radiation four-force. To understand this issue, consider \autoref{eq:G0_2new}, specialising to the case of a grey material, $\chi_0$ independent of $\nu$, for simplicity. In this case, and summing over all groups (or equivalently taking $\nu_{g\pm}\to \pm\infty$), we have
\begin{equation}
    -cG^0 = \chi_0 \left(4\pi B - c E + c^{-1} v^i F^i\right), 
\end{equation}
where $B$, $E$, and $F$ are the Planck function, radiation energy, and radiation flux integrated over all frequencies. Now consider a moving medium that is in thermal equilibrium in its rest frame. The issue that \citeauthor{Lowrie1999}
point out is that, to order $v/c$, the Planck function and radiation energy are the same in the lab and comoving frames, $E = E_0 = 4\pi B_0/c = 4\pi B/c$, but the lab- and comoving-frame fluxes are not the same; the comoving-frame flux $F^i_0 = 0$, but the lab-frame flux is $F^i = (4/3) v^i E$, and this means that $cG^0 = (4/3) \chi_0 v^2 E / c$, which is non-zero even though it should be zero for a medium in equilibrium. To avoid this issue \citeauthor{Lowrie1999} advocate replacing $v^i F^i$ with $v^i F^i_0$ in the expression for $cG^0$. While doing so alleviates the inconsistency, the price of this fix in a multigroup method is that it requires constructing the third moment of the specific intensity, thereby obviating one of the main advantages of our mixed-frame formulation (see \autoref{sec:intro}).

Fortunately, such a fix is also unnecessary, as we now show. It is unnecessary because all that is required for a formulation to be consistent is that in the equilibrium state where $cG^0$ and $G^i$ vanish that $E$ and $\vecF$ take on the correct values, and our formulation \textit{does} satisfy this condition. To see this, note that one can verify by direct substitution into \autoref{eq:G0_2new} and \autoref{eq:Gi_2new} in the case of constant $\chi_0$ that the equilibrium solution that gives $cG^0 = G^i = 0$ is
\begin{equation}
    E = \frac{4\pi B}{c} \left(1 + \frac{4}{3}\frac{v^2}{c^2}\right) \qquad \vecF = \frac{4}{3}\vecv \left(\frac{4\pi B}{c}\right),
\end{equation}
which is in fact the \textit{correct} expression for the lab-frame energy density and radiation flux \citep[c.f.~equation 9.13 of][]{Mihalas1984} to order $v^2/c^2$. The extra $v^2/c^2$ term in $E$ comes from the  $v^i F^i$ and {\it not} from the a higher order expansion in $v/c$ and therefore does not make the formulation more relativistically correct than order $v/c$. Nonetheless, this means that, even though we only have formal accuracy to order $v/c$, our formulation ensures that we recover the correct equilibrium state.

\subsection{Evaluation of group-averaged emissivities and opacities}
\label{sec:RHD2}

\autoref{eq:G0_2new} and \autoref{eq:Gi_2new} involve a series of integrals over quantities that take the form of an opacity multiplied by a radiation quantity. We must therefore now adopt a strategy for evaluating these integrals.
Most previous studies have taken the approach of using either constant opacity within each radiation bin \citep{Zhang2013, Skinner2019}, or some form of weighted opacities, such as the Planck and Rosseland means \citep[e.g.][]{Jiang22a}. The former approach oversimplifies the problem, failing to account for potentially significant variations in opacity within each frequency bin. While the latter is accurate in the diffusion limit, it becomes highly inaccurate outside this regime, where there is no reason to assume that the frequency distribution within a frequency bin resembles a Planck function. To address this, we first implement a simple model with piecewise constant (PC) opacities, assuming the opacity is a constant function of frequency within each radiation group, similar to the approach used in earlier codes. We then introduce a piecewise powerlaw (PPL) approximation, which we will show below offers significantly better accuracy for a very modest increase in computational cost in the common situation where the frequency resolution is low. 

\subsubsection{Piecewise constant opacity}

In the piecewise constant (PC) model, we assume the absorption coefficient is a constant function of $\nu$ within each frequency bin, i.e.,
\begin{equation}\label{eq:m1e1}
\chi_{0}(\nu) = \chi_{0,g},\quad{} {\rm for} \quad{} \numinus < \nu < \nuplus,
\end{equation}
Ignoring the discontinuity at group boundaries, this assumption implies that the $\partial \chi_0 / \partial \nu$ terms in \autoref{eq:G0_2new} and \autoref{eq:Gi_2new} vanish, and by replacing $\chi_{0}(\nu)$ with $\chi_{0,g}$ we obtain
\begin{equation}\label{eq:G_model1b}
    \begin{split}
        -c G_g^0 &= 4 \pi \chi_{0,g} B_g - c \chi_{0,g} E_g + c^{-1} v^i \chi_{0,g} F_g^i \\
        -G_g^i &= -c^{-1} \chi_{0,g} F_g^i + 4 \pi c^{-2} v^j \chi_{0,g} \left(B_g - \frac{1}{3} \Delta_g (\nu B_{\nu}) \right) \\
        & \quad{} + c^{-1} v^j \chi_{0,g} P_g^{ji},
    \end{split}
\end{equation}
where we have introduced the shorthand notation that for any frequency-dependent quantity $Q(\nu)$, we define $\Delta_g(Q) \equiv Q(\nu_{g^+}) - Q(\nu_{g^-})$, i.e., $\Delta_g(Q)$ is simply the difference between $Q(\nu)$ evaluated at the upper and lower frequency limits for a given group. 

Compared to the frequency-integrated formulation \autoref{eq:Gsingle1} and \autoref{eq:Gsingle2}, we notice an extra term that is proportional to $\Delta_g (\nu B_{\nu})$. This term can be calculated analytically at a given gas temperature for each group, and the terms have the property that they vanish when summed over all groups to produce the frequency-integrated four-force $(G^0,\vecG)$ (as long as our frequency grid is broad enough that $\nu B_\nu$ is negligible at both the low- and high-frequency edges of the grid). To understand the physical meaning of these extra terms, recall that $\vecG$ is the total rate of momentum transfer from matter to radiation, and the term $-4\pi c^{-2} v^j \chi_{0P} B$ appearing in \autoref{eq:Gsingle2} represents the momentum transferred from gas to radiation because moving matter produces Doppler-shifted emission that, as a result of the shift, carries a non-zero net momentum. This Doppler shift, however, also changes the distribution of this momentum over frequency, so that the distribution of momentum in frequency is not identical to the distribution of energy. It is this additional difference between the energy and momentum distributions of the thermal emission that is captured by the terms proportional to $\Delta_g(\nu B_{\nu})$.

\subsubsection{Piecewise powerlaw opacity}

While the piecewise constant approach has the advantage of simplicity, and has been the most common approximation used in previous multigroup RHD methods \citep[e.g.,][]{Vaytet2011, Zhang2013, Skinner2019}, it ignores the potentially large variations of the opacity within an energy group that can occur when the frequency resolution is limited, as is often the case in real applications. To better capture this situation we introduce the piecewise powerlaw (PPL) approximation; this approach is somewhat similar to the one proposed by \citet{Hopkins2023c}, but here we extend this method to the mixed-frame formulation of RHD, including the Doppler effect terms responsible for energy exchange between energy groups. 

Our fundamental \textit{ansatz} is to assume the absorption coefficients can be expressed as powerlaw functions of frequency over each bin: 
\begin{equation}
\chi_{0g}(\nu) = \chi_{0,g-} \left(\frac{\nu}{\numinus}\right)^{\alpha_{\chi_0,g}},\qquad\numinus \le \nu \le \nuplus.
\end{equation}
We note that part of our motivation for this \textit{ansatz} is that powerlaw functional forms are often expected on physical grounds; for example, both infrared dust opacities and free-free opacities are close to powerlaws in frequency. Thus in many cases we know the bin edges opacities $\chi_{0,g-}$ and powerlaw indices $\alpha_{\chi_0,g}$ simply from the physical nature of the opacity. In cases where this is not the case and the opacity $\chi_0$ is tabulated, we can adopt the approximation
\begin{equation}
    \alpha_{\chi_0,g} = \frac{\ln \left[\chi_0(\nuplus)/\chi_0(\numinus) \right]}{\ln (\nuplus/\numinus)}.
\end{equation}
Our implementation in \textsc{quokka} allows users to either use this approach or specify $\alpha_{\chi_0,g}$ for each group directly, covering either case. 

Inserting this functional form into \autoref{eq:G0_2new} and \autoref{eq:Gi_2new} we obtain for the time-like component of the radiation four-force 
\begin{equation}\label{eq:G0_3}
    \begin{split}
    - c G_{g}^0 &= 4 \pi \chi_{0B,g} B_g - c \chi_{0E,g} E_g \\ 
    & \quad{} + c^{-1} (1 + \alpha_{\chi_0,g}) v^i \chi_{0F,g}^{(i)} F_g^i, 
    \end{split}
\end{equation}
and for the space-like component 
\begin{equation}\label{eq:G1_3}
    \begin{split}
    - G_{g}^i &= - c^{-1} \chi_{0F,g}^{(i)} F_g^i \\
    & \quad{} + 4 \pi c^{-2} v^i \left[ \chi_{0B,g} B_g -\frac{1}{3} \Delta_g (\nu \chi_0 B_{\nu}) \right] \\
    & \quad{} + c^{-1} v^j (1 + \alpha_{\chi_0,g} ) \chi_{0P,g}^{(ji)} P_g^{ji},
    \end{split}
\end{equation}
where 
\begin{equation}\label{eq:chiQ}
    \chi_{0Q,g} \equiv \frac{\intg \chi_{0}(\nu) Q_{\nu}(\nu) d\nu}{Q_g}
\end{equation}
is the mean opacity at group $g$ averaged over a spectrum $Q_{\nu}(\nu)$, where $Q$ can be $B, E, F^i$, or $P^{ji}$. The opacity with a superscript $i$ or $ji$ represent the opacity weighted by the $i$ or $ji$ components of the corresponding radiation quantity. For reasons of speed and code simplicity we further assume that $\chi_{0P,g}^{(ji)} \approx \chi_{0E,g}$, which is equivalent to neglecting the frequency-dependence of the Eddington tensor when evaluating mean opacities.

\paragraph{Relations between the mean opacities}

We next note that the mean opacities $\chi_{0E,g}, \chi_{0B,g}$, and $\chi_{0F,g}$ appearing in \autoref{eq:G1_3} are required to obey certain relations in the limit of high optical depth if we are to ensure that our method limits to the correct diffusion solution, one of the important design goals of our algorithm. The obvious requirement for this to be the case is that at high optical depth the radiation energy spectrum $E_\nu$ approaches the Planck function $B_\nu$, and thus to ensure that $G^0_g$ vanishes in this limit, we require that $\chi_{0E,g}$ and $\chi_{0B,g}$ approach the same value at high optical depth. We can guarantee that this is the case simply by using the same method to estimate both quantities.

The more subtle requirement involves $\chi_{0F,g}$. For an optically thick, uniform medium at rest we should have $F_\nu \to 0$, in which case our choice of $\chi_{0F,g}$ does not matter. But now consider a moving homogeneous medium, for which $F_\nu$ is non-zero. Indeed, examining the spacelike component of the radiation four-force, we see that the condition for radiation-matter equilibrium (i.e., $G_{\nu}^i \to 0$) reduces to
\begin{equation}
\begin{split}
    F_{\nu}^i &\to \frac{4}{3} \pi c^{-1} v^i \left( 2 B_{\nu} - \frac{\partial \ln \chi_0}{\partial \ln \nu} B_{\nu} - \frac{\partial B_{\nu}}{\partial \ln \nu} \right) \\
    & \quad {} + \left( 1 + \frac{\partial \ln \chi_0}{\partial \ln \nu} \right) v^j P_{\nu}^{ji}.
\end{split}
\end{equation}
In the optically thick limit the radiation pressure tensor (to order $v/c$) is $\tenP_{\nu} = \frac{1}{3} \frac{4 \pi B_{\nu}}{c} \tenI$, and we obtain 
\begin{equation}
\begin{split}
    F_{\nu, \rm diff}^i &= 4 \pi c^{-1} v^i \left( B_{\nu} - \frac{1}{3} \frac{\partial B_{\nu}}{\partial \ln \nu} \right).
\end{split}
\end{equation}
Then, we integrate $F_{\nu, \rm diff}^i$ within a energy bin and get the group-integrated flux in the diffusion limit
\begin{equation}\label{eq:Fdiff}
    F_{g, \rm diff}^i = 4 \pi c^{-1} v^i \left[ \frac{4}{3} B_g - \frac{1}{3} \Delta_g (\nu B_{\nu}) \right].
\end{equation}
If we insert this result into \autoref{eq:G1_3}, we see that in order to guarantee that our numerical approximation to $G^i_g$ properly approaches zero at high optical depth in a moving medium, our expression for the group-averaged flux-mean opacity must approach 
\begin{multline}\label{eq:chiDiff}
        \chi_{0F,g,\rm diff} = \\
        \frac{
        (\chi_{0B,g} + \frac{1}{3} \chi_{0E,g}) B_g + \frac{1}{3} \left[ \alpha_{\chi_0,g} \chi_{0E,g} B_g - \Delta_g (\nu \chi_0 B_{\nu} ) \right]
        }{
        \frac{4}{3} B_g - \frac{1}{3} \Delta_g (\nu B_{\nu})
        }.
\end{multline}
Note that, as we might expect, \autoref{eq:chiDiff} reduces to $\chi_{0F,g,\rm diff} = \chi_{0,g}$ in the PC opacity case where $\chi_{0B,g} = \chi_{0E,g} = \chi_{0,g}$. In principle any method of approximating $\chi_{0F,g}^{(i)}$ with the property that it approaches $\chi_{0F,g,\rm diff}$ at high optical depth would satisfy this requirement. However, for simplicity we choose to enforce this requirement in all cases and in all directions, so that we have completely specified $\chi_{0F,g}^{(i)}$ in terms of $\chi_{0E,g}$, $\chi_{0B,g}$, and the matter properties. 

\paragraph{PPL approximations to radiation quantites}

We have now reduced our problem to the selection of an algorithm to estimate $\chi_{0E,g}$ and $\chi_{0B,g}$. For the latter, since $B_\nu$ is a function only of the matter temperature, we could of course evaluate the average directly. However, adopting this method would carry a significant disadvantage: since we have just shown that we must use the same algorithm to evaluate $\chi_{0E,g}$ and $\chi_{0B,g}$, evaluating $\chi_{0B,g}$ directly from the matter temperature would force us do the same for $\chi_{0,Eg}$, which would amount to assuming that the radiation spectrum within each bin looks like a blackbody spectrum. While this is a less restrictive assumption than it would be a grey method, since we would be assuming a thermal spectrum only within each frequency bin rather than across all frequencies, we nonetheless wish to avoid making it because it is highly inaccurate at low optical depth.

Instead, we consider an alternative approach whereby we assume that both $E_\nu$ and $B_\nu$ can be expressed approximately as powerlaw functions of frequency within each bin,
\begin{equation}
Q_{\nu}(\nu) = Q_{\numinus} \left(\frac{\nu}{\numinus}\right)^{\alpha_{Q,g}},\qquad\numinus \le \nu \le \nuplus, 
\end{equation}
where $Q_{\numinus}$ is chosen such that 
\begin{equation}
    \intg Q_{\nu}(\nu) d\nu = Q_g.
\end{equation}
As with our PPL limit for the opacities, this choice is motivated by the fact that for many physical situations radiation frequency distributions are in fact well described by powerlaw over large ranges in frequency; an obvious example of this is the Rayleigh-Jeans tail of the Planck function. For this choice one can show that \autoref{eq:chiQ} evaluates to
\begin{equation}
    \label{eq:chiQg}
    \chi_{0Q,g} = \chi_{0,g-} \left[ \frac{ r_g^{\alpha_{Q,g} + 1} - 1}{\alpha_{Q,g} + 1} \right]^{-1} {} \left[ \frac{r_g^{\alpha_{\chi_0,g} + \alpha_{Q,g} + 1} - 1}{\alpha_{\chi_0,g} + \alpha_{Q,g} + 1} \right]
\end{equation}
where for convenience we have defined $r_g \equiv \nuplus/\numinus$; note that the first and second terms in square brackets are replaced by $\ln r_g$ for the special cases $\alpha_{Q,g} = -1$ and $\alpha_{\chi_0,g} + \alpha_{Q,g} = -1$, respectively. Also note that, as one might expect, the expressions above reduce to the PC case for $\alpha_{\chi_0,g} = 0$, and that $\alpha_{Q,g}$ matters only when $\alpha_{\chi_0,g} \neq 0$, i.e., we care about the frequency-dependence of radiation quantities only if the material opacity varies across a frequency bin.

The remaining question is how to choose the indices $\alpha_{Q,g}$. For this purpose we consider two possible methods. One is to determine $\alpha_{0B,g}$ and $\alpha_{0E,g}$ directly by fitting $E_\nu$ and $B_\nu$ to a PPL functional form, and we describe a method to do so in \aref{sec:slope}. However, we show below that this approach does not perform any better than the favoured one we describe next, and has a significantly higher computational cost. Thus while we include it for testing purposes, in practice we do not recommend using it.

After testing several approaches, our favoured one is to make the simple assumption that $\nu Q_\nu$ is constant across a bin, or equivalently that $\alpha_{Q,g} = -1$; we refer to this as the PPL fixed-slope method. In addition to having the advantage of simplicity and therefore very low computational cost, this approximation gives the correct \textit{average} value of $\alpha_{Q,g}$, which we can see by noting that this average for any quantity $Q(\nu)$ is given by
\begin{equation}\label{eq:equality}
    \begin{split}
        \frac{\int_0^{\infty} \alpha_{Q}(\nu) Q(\nu) \, d \nu}{\int_0^\infty Q(\nu)\, d\nu} &= \frac{\int_0^{\infty} \frac{d \ln Q(\nu)}{d \ln \nu} Q(\nu) \,d \nu}{\int_0^\infty Q(\nu)\, d\nu} \\
        &= \frac{\int_{\nu=0}^{\nu=\infty} \nu \,d Q(\nu)}{\int_0^\infty Q(\nu)\, d\nu} \\
        &= \frac{\left. \nu Q(\nu) \right|_{\nu=0}^{\nu=\infty} - \int_0^{\infty} Q(\nu) d \nu}{\int_0^\infty Q(\nu)\, d\nu} \\
        &= -1,
    \end{split}
\end{equation}
where in the final step we note that $\nu B_\nu\to 0$ at $\nu = 0$ and $\infty$ for any finite matter temperature, and that $\nu E_\nu$ must similarly approach zero at $\nu = 0$ and $\infty$ in order to ensure that the total frequency-integrated radiation energy is finite. Thus the mean powerlaw index of both $E_\nu$ and $B_\nu$, when weighted by the spectrum itself, is $-1$.  %
Our numerical tests have demonstrated that this PPL fixed-slope method performs effectively in handling problems with rapidly varying opacity across frequencies, even when using a small number of energy bins (see \autoref{sec:marshak2} and \autoref{sec:pulse}). 

\section{Numerical method}
\label{sec:numerical}

The overall structure of a time step for our multigroup method follows the same asymptotic-preserving structure that we introduced for grey radiation in \citetalias{He2024}, which we only briefly summarise here. We solve the system formed by \autoref{eq:hyper} with state vector, transport terms, and source terms given by \autoref{eq:MG_state_vec} using an operator split approach. In the first step we solve the hydrodynamic transport subsystem 
\begin{equation}
    \frac{\partial}{\partial t} \left[
    \begin{array}{c}
      \rho \\
      \rho \boldsymbol{v} \\
      E_{\rm gas}
    \end{array}\right] 
    + \nabla \cdot \left[
    \begin{array}{c}
      \rho \boldsymbol{v} \\
      \rho \boldsymbol{v} \otimes \boldsymbol{v}+p \\
      (E_{\rm gas} + p) \boldsymbol{v}
    \end{array}\right]
    = 0
\end{equation}
using a method of our choice; for all the tests presented below the method we use is as described in \citetalias{Wibking2022}. In the second step we update the radiation transport and matter-radiation exchange subsystems, using time steps that are sub-cycled with respect to the hydrodynamic update. Following \citetalias{He2024}, we handle these subsystems using a method of lines formulation where we define $E_{ijk}$ as the volume-averaged radiation energy density in cell $ijk$, and similarly for all other variables, and express the radiation subsystem for each cell as 
\begin{equation} \label{eq:dysf}
    \frac{d}{dt} \matU_{ijk} = \matT(\matU_{ijk}, t) + \matS(\matU_{ijk}, t),
\end{equation}
where, dropping the $ijk$ subscript from this point forward for convenience and introducing the reduced-speed-of-light approximation, we define 
\begin{equation} \label{eq:ysf}
  \matU = \left[
    \begin{array}{c}
      \rho \boldsymbol{v} \\
      E_{\rm gas} \\
      E_g \\
      \frac{1}{c \chat} \boldsymbol{F}_g
    \end{array}\right], \quad
    \mathbfss{T} = - \nabla \cdot \left[
    \begin{array}{c}
      0 \\
      0 \\
      \frac{\chat}{c} \boldsymbol{F}_g \\
      \tenP_g
    \end{array}\right], \quad 
    \matS =
    \left[
    \begin{array}{c}
      \sum_g \boldsymbol{G}_g \\
      c \sum_g G_{g}^0 \\
      - \chat G_{g}^0 \\
      - \boldsymbol{G}_g
    \end{array}\right],
\end{equation}
as the vectors of conserved quantities, the transport terms, and the source terms for the radiation subsystem update. In these equations $\hat{c}$ is the reduced speed of light, which we can set to value value $<c$ to reduce time step constraints. We provide this formulation for completeness, but unless otherwise noted below we always set $\hat{c} = c$, thereby recovering the exact equations. We solve this system using the IMEX method introduced in \citetalias{He2024} to integrate \autoref{eq:dysf} over a time $\Delta t$ in a two-stage process:
\begin{align}
    \matU^{(n+1/2)} =& ~\matU^{(n)} + \Delta t ~\matT (\matU^{(n)}) + \Delta t ~\matS (\matU^{(n+1/2)}) \label{eq:imex1} \\
    \matU^{(n+1)} =& ~\matU^{(n)} + \Delta t \left[ \frac{1}{2} ~\matT(\matU^{(n)}) + \frac{1}{2} ~\matT(\matU^{(n+1/2)}) \right] \nonumber \\
    &+ \Delta t \left[ \frac{1}{2} ~\matS(\matU^{(n+1/2)}) + \frac{1}{2} ~\matS(\matU^{(n+1)}) \right], \label{eq:imex2}
\end{align}
where the superscript $(n)$ indicates the state at the start of the radiation update (but after the operator-split hydrodynamic update), $(n+1)$ indicates the state at the end of the radiation update, and $(n+1/2)$ indicates an intermediate stage. As discussed in \citetalias{He2024}, this scheme has the advantage of asymptotically preserving the diffusion limit and requiring no more communication than hydrodynamics, which is a huge advantage on modern GPU-based structures where communication is expensive. We refer the reader to \citetalias{He2024} for details of the time-integration scheme. 

Evaluating each stage of this two-stage integration requires first evaluating the explicit terms -- those that depend only on $(n)$ quantities at the first stage, and those that depend on $(n)$ or $(n+1/2)$ quantities at the second stage -- and then solving the remaining implicit equation for the next stage terms -- $(n+1/2)$ at the first stage and $(n+1)$ at the second stage. The extension of the explicit terms to the multigroup case is trivial, since one can simply evaluate them group-by-group. We therefore focus on the implicit stage, which requires modification for the multigroup case. To start with, for both stages of the IMEX integrator we can express the implicit equation to be solved in the generic form 
\begin{gather}\label{eq:egupdate}
    E_{\rm gas}^{(t+1)} - E_{\rm gas}^{(t)} = c \theta \Delta t \sum_g G_{g}^{0,(t+1)} \\
    \label{eq:eupdate}
    E^{(t+1)}_g - E^{(t)}_g = - \chat \theta \Delta t G_{g}^{0,(t+1)} \quad {\rm for} \ g = 1, 2, \cdots, N
\end{gather}
for the energies, and 
\begin{gather}
(\rho \vecv)^{(t+1)} - (\rho \vecv)^{(t)} = \theta\Delta t \sum_g \vecG^{(t+1)}_g \label{eq:vupdate} \\
\vecF^{(t+1)}_g - \vecF^{(t)}_g = - \chat c \theta\Delta t \vecG^{(t+1)}_g \quad {\rm for} \ g = 1, 2, \cdots, N_g \label{eq:fupdate}
\end{gather}
for the momenta. Here terms with superscript \((t)\) denote the state after applying the explicit terms -- for example during the first stage (\autoref{eq:imex1}) we have $E^{(t)}_g = E^{(n)}_g - \Delta t (\hat{c}/c) \nabla\cdot \vecF^{(n)}_g$ -- and terms with superscript \((t+1)\) denote the final state for which we are attempting to solve; the factor $\theta$ is unity during the first stage and $1/2$ during the second stage. 

In each cell this is a system of  $4 + 4N_g$ equations in $4 + 4 N_g$ unknowns -- gas energy and $N_g$ radiation energies, three components of gas velocity, and three components of $N_g$ radiation fluxes -- that must be solved simultaneously. As discussed in \citetalias{He2024}, it is numerically convenient to divide this problem into an inner and an outer stage; in the inner stage we solve the energy equations while freezing $\vecv$ and $\vecF_g$, and in the outer stage we update the fluxes and gas velocity and then if necessary go back to the inner stage and recompute $E_g$ and $E_{\rm gas}$ using the updated values of $\vecv$ and $\vecF_g$. The reason this is more efficient is that it allows the inner iteration stage to consider only $1 + N_g$ variables rather than $4 + 4 N_g$, making it faster, and in most cases the work terms proportional to $\vecv$ and $\vecF_g$ are small in the energy equations, so $E_g$ and $E_{\rm gas}$ change little to none as a result of the update to $\vecv$ and $\vecF_g$ and the whole procedure converges in a single or at most a few outer iterations. 

The inner stage consists of solving \autoref{eq:egupdate} and \autoref{eq:eupdate} for the new energies, which we do via Newton-Raphson iteration following \citet{Howell2003},  \citetalias{Wibking2022}, and \citetalias{He2024}. For the multi-group case, we can write out the system to be solved as 
\begin{equation} \label{eq:FGFR}
\left\{
\begin{split}
    0 &= F_G(E_{\rm gas}^{(t+1)}, E^{(t+1)}_1, E^{(t+1)}_2, \cdots) \\
    &\equiv E_{\rm gas}^{(t+1)} - E_{\rm gas}^{(t)} + \left( \frac{c}{\chat} \right) \sum_g R_g^{(t+1)} \\
    0 &= F_{R,1}(E_{\rm gas}^{(t+1)}, E_1^{(t+1)}, E_2^{(t+1)}, \cdots) \\
    &\equiv E_1^{(t+1)} - E_1^{(t)} - (R_1^{(t+1)} + S_1^{(t+1)}), \\
    \vdots \\
    0 &= F_{R,N_g}(E_{\rm gas}^{(t+1)}, E_1^{(t+1)}, E_2^{(t+1)}, \cdots) \\ 
    &\equiv E_{N_g}^{(t+1)} - E_{N_g}^{(t)} - (R_{N_g}^{(t+1)} + S_{N_g}^{(t+1)}),
\end{split}
\right.
\end{equation}
where 
\begin{equation} \label{eq:Rnew}
    \begin{split} 
    R_g^{(t+1)} &\equiv - \chat G_{g}^{0,(t+1)} \theta \Delta t \\
    &= \left[ 4 \pi \chi_{0B,g} B_g - c \chi_{0E,g} E_g \right. \\ 
    & \quad{} \left. + c^{-1} (1 + \alpha_{\chi_0,g}) v^i \chi_{0F,g}^{(i)} F_g^i \right] \theta \Delta t 
    \end{split}
\end{equation}
and $S_g$ is an optional term to include, for example, the addition of radiation by stellar sources. We remind readers here that, although we have omitted them to avoid clutter, $E_g$ and all the variables that can depend on $E_{\rm gas}$ -- $\chi_{0B,g}$, $\chi_{0E,g}$, $\chi_{0F,g}$, and $B_g$ -- carry the superscripts ${(t+1)}$ to indicate that they are evaluated at the new time, while $\vecv$ and $\vecF_g$ are fixed to their values at the start of the inner stage as noted above, and thus do not carry subscripts $(t+1)$ and do not evolve during the inner Newton-Raphson stage.

A single Newton-Raphson iteration for this system consists of solving the linearized equations 
\begin{equation}
  \label{eq:linear}
  \mathbf{J} \cdot \Delta \vecx = - \vecF(\vecx),
\end{equation}
where $\vecx$ is the set of variables to be updated, $\Delta\vecx$ is the change in these variables during this iteration, $\vecF(\vecx)$ is the vector whose zero we wish to find, and $\mathbf{J}$ is the Jacobian matrix of \( \vecF(\vecx) \). As discussed in \citetalias{He2024}, we use the gas energy and the source terms as the base variables over which to iterate and compute the Jacobian: $\vecx = (E_{\rm gas}, R_1, R_2, \cdots, R_{N_g})$. The Jacobian in this basis is
\begin{align}
    \frac{\partial F_G}{\partial E_{\rm gas}} &= 1 \\
    \frac{\partial F_G}{\partial R_g} &= \frac{c}{\hat{c}} \\
    \frac{\partial F_{R,g}}{\partial E_{\rm gas}} &= \frac{\partial E_{g}}{\partial E_{\rm gas}} |_{R_g = {\rm const.}} = \frac{1}{C_v} \frac{\partial}{\partial T} \left( \frac{\chi_{0B,g}}{\chi_{0E,g}} \frac{4 \pi B_g}{c} \right) \\
    \frac{\partial F_{R,g}}{\partial R_g} &= \frac{\partial E_{g}}{\partial R_g}|_{T = {\rm const.}} - 1 =  - \frac{1}{\hat{c} \ \theta \Delta t \ \chi_{0E,g}} - 1
\end{align}
In practice, we assume $\partial (\chi_{0B,g}/\chi_{0E,g})/\partial T = 0$ in the calculation of the Jacobian for simplicity, but this simplification only changes the rate of convergence; it does not affect the converged solution. 

The Jacobian matrix is sparse:
\begin{equation}\label{eq:matrix}
  \mathbf{J} =
  \left[\begin{array}{ccccc}
          \frac{\partial F_G}{\partial E_{\rm gas}} & \frac{\partial F_G}{\partial R_1} & \frac{\partial F_G}{\partial R_2} & \cdots & \frac{\partial F_G}{\partial R_{N_g}} \\
          \frac{\partial F_{R,1}}{\partial E_{\rm gas}} & \frac{\partial F_{R,1}}{\partial R_1} & 0 & \cdots & 0 \\
          \frac{\partial F_{R,2}}{\partial E_{\rm gas}} & 0 & \frac{\partial F_{R,2}}{\partial R_2} & \cdots & 0 \\
          \vdots & \vdots & \vdots & \ddots & \cdots \\
          \frac{\partial F_{R,N_g}}{\partial E_{\rm gas}} & 0 & 0 & \cdots & \frac{\partial F_{R,N_g}}{\partial R_{N_g}}.
    \end{array}\right]
\end{equation}
Physically, this sparse structure is a manifestation of the fact that the different energy groups are coupled only via the gas, not directly with each other. This allows inversion of the matrix via
Gauss-Jordan elimination rather than via more general, and much more expensive, matrix inversion methods. First, we cancel the 2nd, 3rd, 4th, ..., elements of the first row with the 2nd, 3rd, 4th, ..., rows, respectively, which eliminates all variables except $\Delta x_1$ and allows us to solve for it. Then, we substitute $\Delta x_1$ into the 2nd row and solve for $\Delta x_2$, and substitute $\Delta x_1$ into the 3rd row and solve for $\Delta x_3$, and so forth. These steps take only $O(N_g)$ floating-point operations, compared to $O(N_g^3)$ operations from a general Gaussian elimination algorithm. 

After solving \autoref{eq:linear} for $\Delta \vecx$, we update $\vecx \leftarrow \vecx + \Delta \vecx$; we then repeat the procedure of solving for $\Delta \vecx$ and updating $\vecx$ until the system converges, as determined by the condition
\begin{equation}
    \frac{\left| F_G \right|}{E_{\rm tot}} < \epsilon \quad {\rm and} \quad \frac{c}{\chat} \frac{\sum_g \left| F_{R,g} \right|}{E_{\rm tot}} < \epsilon
\end{equation}
where 
\begin{equation}
    E_{\rm tot} = E_{\rm gas}^{(t)} + \frac{\chat}{c} \left(\sum_g E_g^{(t)} + \sum_g S_g^{(t)}\right).
\end{equation}
is the total radiation and material energy at the beginning of the timestep accounting for the reduced speed of light. We set the relative tolerance $\epsilon = 10^{-11}$ by default. Once this Newton-Raphson system converges, we have solved for $E_{\rm gas}^{(t+1)}, R_0^{(t+1)}, R_1^{(t+1)}, \cdots$. We update the radiation energy with $E_{g}^{(t+1)} = E_{g}^{(t)} + R_{g}^{(t+1)} + S_{g}^{(t+1)}$ and update the material energy with $E_{\rm gas}^{(t+1)} = E_{\rm gas}^{(t)} - (c/\hat{c}) \sum_g R_g^{(t+1)}$. Note that, for $\hat{c} = c$ and $S_g = 0$, this guarantees conservation of total energy to machine precision regardless of the level of accuracy with which we have iterated the equations to convergence. 

We then proceed to the outer stage of the iteration where we solve the flux and momentum update equations, \autoref{eq:vupdate} and \autoref{eq:fupdate}, with the updated gas temperature, opacity, and radiation energy. For simplicity, we use the outdated velocity while updating radiation flux. The solution to \autoref{eq:fupdate} is straightforward: 
\begin{equation}\label{eq:numF1}
\begin{split}
    F_g^{i,(t+1)} &= \left\{ F_g^{i,(t)} + \hat{c} \theta \Delta t \left[ 4 \pi c^{-1} v^i \left(\chi_{0B,g} B_g - \frac{1}{3} \Delta_g (\nu \chi_0 B_{\nu})\right) \right. \right. \\
    & \quad{} \quad{} + \left. \left. v^j (1 + \alpha_{\chi_0,g}) \chi_{0E,g} P_g^{ji} \right] \right\} / \left( 1 + \chat \chi_{0F,g} \theta\,\Delta t \right).
\end{split}
\end{equation}
Lastly, we update the gas momentum via 
\begin{equation}
    (\rho v^i)^{(t+1)} = (\rho v^i)^{(t)} - (c \chat)^{-1} \sum_g \left( F_g^{i,(t+1)} - F_g^{i,(t)} \right).
\end{equation}
This update also ensures momentum conservation to machine precision. After we update the gas momentum, we also recalculate the gas's internal energy (which in \quokka{} we track separately because we implement a dual energy formalism) by subtracting the updated kinetic energy from the updated total energy. 

As previously indicated, the gas velocity $\vecv$ and radiation flux $\vecF$ we use in the inner stage of the iteration are lagged. This can cause significant inaccuracies at high optical depths when the velocity-dependent terms are non-negligible. To eliminate errors like this and render this scheme fully implicit, we now repeat the inner iteration using the updated values of $\vecv$ and $\vecF$, and compute new estimates for $\vecv$ and $\vecF$, repeating this procedure until the sum of the absolute change in the value of the terms proportional to $\vecv$ and $\vecF$ in $R$ (\autoref{eq:Rnew}) from one outer iteration to the next is below $10^{-13} \sum_g R_{g}$ and $10^{-13} E_{\rm tot}$. Except in the dynamic diffusion limit, where the velocity-dependent terms are at the same order as all other terms, this iterative process typically terminates after just one iteration, and in all the tests we present below, and for all the test problems presented in \citetalias{Wibking2022} and \citetalias{He2024}, we never require more than a handful of outer iterations. Thus the cost is modest. This completes accounting for all terms in the radiation four-force, thus completing radiation-matter coupling.

\section{Numerical Tests}
\label{sec:tests}

We next carry out a series of tests of our numerical method to verify its accuracy. Our tests can be divided into those that purely test the radiative transfer part of the code (\autoref{ssec:rad_tests}) and those that test the full radiation-hydrodynamics functionality (\autoref{ssec:rhd_tests}).

\subsection{Radiation transport tests}
\label{ssec:rad_tests}

We begin with tests in which we disable  the hydrodynamic update parts of the code, and we test the radiation transport parts in isolation.

\subsubsection{Multigroup Marshak Waves}\label{sec:marshak}

\begin{figure*}
    \centering
    \includegraphics[]{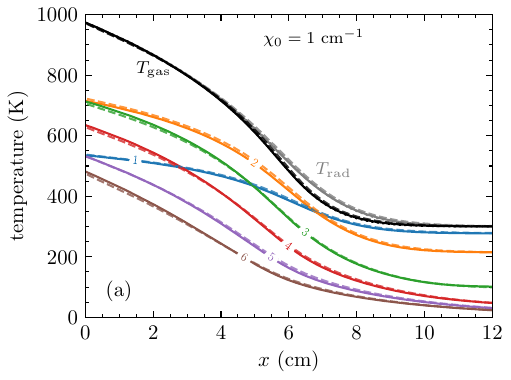}\includegraphics[]{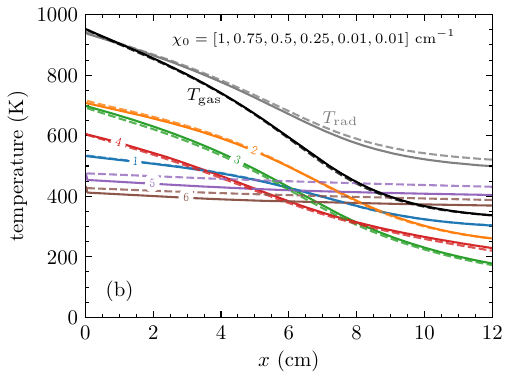}\\
    \includegraphics[]{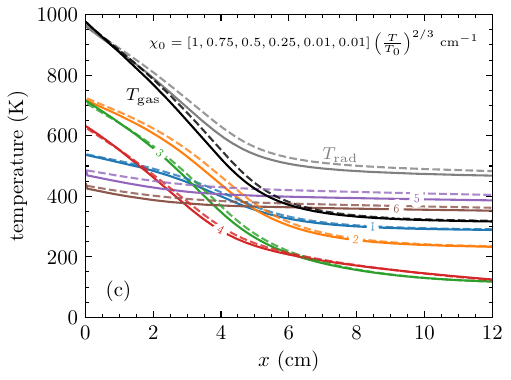}
    \caption{Multigroup Marshak wave tests calculated using three different opacity models: constant opacity (panel a), frequency-dependent opacity (panel b), and frequency- and temperature-dependent opacity (panel c). We plot the gas temperature, marked with $T_{\rm gas}$, the radiation temperature, marked with $T_{\rm rad}$, and the temperature of 6 radiation groups, marked with the group index; see main text for our definitions of $T_\mathrm{rad}$ and the group radiation temperatures. We compare the numerical solutions (solid lines) computed using \textsc{quokka} with exact ray-tracing solutions (dashed lines, data adapted from \citealt{Vaytet2011} through personal communication) and find good agreement.}
    \label{fig:marshak}
\end{figure*}

Our first test, taken from \citet{Vaytet2011}, consists of simulations of Marshak waves. We consider three different models for opacity: (a) constant opacity, (b) frequency-dependent opacity, and (c) frequency- and temperature-dependent opacity. The goal is to evaluate the accuracy of the multigroup method in handling variable opacities and to study how these opacities affect wave propagation. 

In all three simulations, the gas is at rest with a uniform density $\rho = 10^{-3} \ {\rm g~cm^{-3}}$ and a temperature $T = 300$ K, in equilibrium with the radiation. The simulations use six frequency groups, with the first five groups evenly spaced between $\nu=0$ and $\nu = 1.5 \times 10^{14}$ Hz, the last group covering the range from $1.5 \times 10^{14}$ Hz to $\infty$. The heat capacity of the gas is set to $\rho C_V = 10^{-3} \ {\rm erg~cm^{-3}~K^{-1}}$. The 1D spatial grid extends from 0 to 20 cm, resolved using 500 cells. The boundary conditions on the radiation are set to zero flux initially, with an energy density in each group corresponding to blackbody radiation at temperatures of $1000$ K and $300$ K on the left and right sides of the domain, respectively. We run the simulations to $t = 1.36 \times 10^{-7}$ s. 

In simulation (a), the gas has a constant specific opacity $\kappa_0 = 1000 \ {\rm cm^{2}~g^{-1}}$ independent of frequency. The total opacity is $\chi_0 = \rho \kappa_0 = 1 \ {\rm cm^{-1}}$, so in this simulation the optical depth across the simulation domain is 20. For the frequency-independent material opacity the PC and PPL methods are identical, so we need not distinguish them. We show the results from this simulation in the first panel of \autoref{fig:marshak}, where the curves represent the gas temperature, labeled as $T_{\rm gas}$, the radiation temperature, labeled as $T_{\rm rad}$, and the radiation temperature inside each individual energy group $T_{\mathrm{rad},g}$, labeled as an integer $g$. We define $T_{\mathrm{rad},g}$ by the following relation:
\begin{equation}
    \label{eq:Tgdef}
    E_g = a_R T_{\mathrm{rad},g}^4,
\end{equation}
and we define the total radiative temperature $T_\mathrm{rad}$ by
\begin{equation}
    \sum_g E_g = a_R T_\mathrm{rad}^4. 
\end{equation}
The numerical calculation is performed on the domain [0, 20] cm, but we limit the $x$ axis to [0, 12] cm, which is the most relevant portion of the data.
We compare $T_{\rm gas}$, $T_{\rm rad}$, and $T_{\mathrm{rad},g}$ from our numerical calculation (solid lines) with the exact ray-tracing solution (dashed lines) from \cite{Vaytet2011}, finding excellent agreement. In particular, we see that we reproduce an important feature in the solution: although the opacity is the same in all groups, and thus all groups have the same diffusion speed, variation in the overall radiation temperature across the domain leads to a shift in which frequency groups dominate, with group 1 dominating in the upstream, cool gas and groups 2, 3, and 4 becoming dominant in the heated regions. In addition to demonstrating the overall accuracy of our method, this test demonstrates the validity of the $M_1$ closure approximation in our code, at least for this problem.

We also carry out one additional comparison here, by running the problem using the grey method described in \citetalias{He2024} -- for frequency-independent opacity, this method is highly accurate. We compare $T_{\rm gas}$ and $T_{\rm rad}$ computed the multigroup model presented in this paper (solid lines) to those computed using the grey method (dotted line), and find excellent agreement -- the lines for multigroup and single-group models are indistinguishable to the eye. This demonstrates that the multigroup scheme consistently reduces to a grey model for frequency-independent opacities. 

Simulation (b) employs group-specific opacities: $\chi_{0,g} = [1, 0.75, 0.5, 0.25, 0.01, 0.01] \ {\rm cm}^{-1}$ for groups 1-6, respectively; following \citet{Vaytet2011}, the opacities are assumed to be constants across groups, and thus we use the PC opacity model, or equivalently the PPL model with $\alpha_{\chi_0,g} = 0$ for all groups. For these opacities the optical depth of the domain varies from 20 in group 1 to 0.2 in groups 5 and 6. The results are shown in the second panel of \autoref{fig:marshak}. Unlike in \cite{Vaytet2011}, we do not extend the grid to the right with extra cells. Compared to the solution in panel (a), we see that radiation in groups with lower opacities (notably groups 5 and 6) rapidly traverses the entire grid and heats the gas at the right edge, raising the gas temperature to 330 K from its initial value of 300 K. This pre-heating by the more-rapidly propagating radiation in the low-opacity groups in turn causes the the radiation in groups 1 and 2 to advance slightly further than in the constant opacity case. Comparison between the numerical solution and exact ray-tracing solutions is again excellent for the gas temperature. The total radiative temperatures are slightly larger than in simulation (a), particularly in the low-opacity groups 5 and 6. As \cite{Vaytet2011} point out, this is primarily a boundary condition effect: when the difference between the left and right fluxes is large (which is the case for groups 5 and 6 at the left boundary since the flux in the ghost cells is set to zero), the $M_1$ model becomes less accurate. 

Simulation (c) features both frequency- and temperature-dependent opacities: $\chi'_{0,g} = \chi_{0,g} (T/T_0)^{3/2}$, where $T_0 = 300$ K and $\chi_{0,g}$ in the groups 1-6 are the same as in simulation (b), and we again use PC opacities to be consistent with the assumptions used to derive the reference solution. The results are shown in the third panel of the figure. As the gas temperature $T_\mathrm{gas}$ increases, so does the opacity, leading to more rapid radiation absorption. Group 1 is less affected by this effect due to re-radiation from the hot gas. We once again see good agreement between the multigroup $M_1$ model and the exact ray-tracing model, especially for the gas temperature. 

Overall, these Marshak wave simulations demonstrate the capability of the multigroup method to handle variable opacities accurately and highlight the effects of different opacity models on Marshak wave propagation.

\subsubsection{Marshak waves with continuously frequency-dependent opacity: convergence rates for different opacity models}\label{sec:marshak2}

\begin{figure*}
    \centering
    \includegraphics[width=0.5\textwidth]{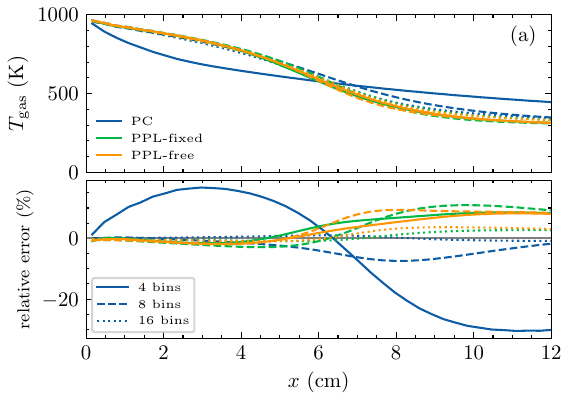}\includegraphics[width=0.5\textwidth]{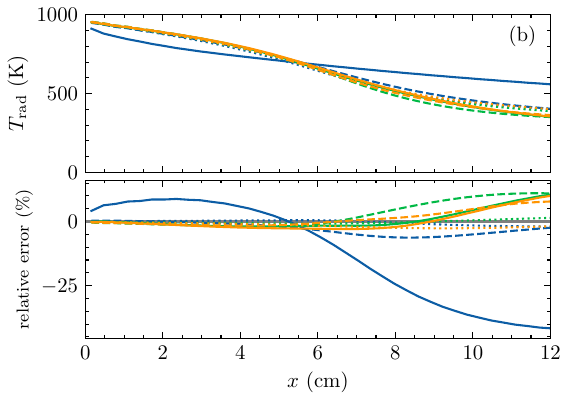}\\
    \includegraphics[width=0.5\textwidth]{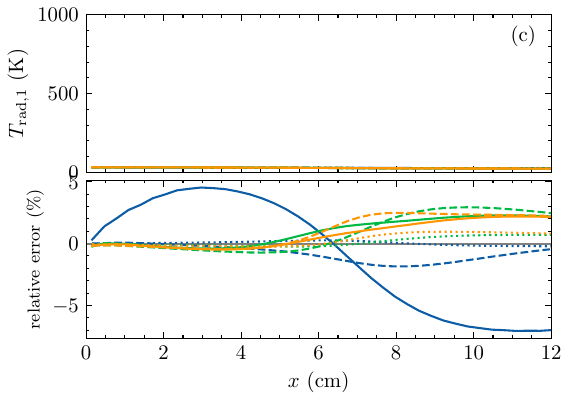}\includegraphics[width=0.5\textwidth]{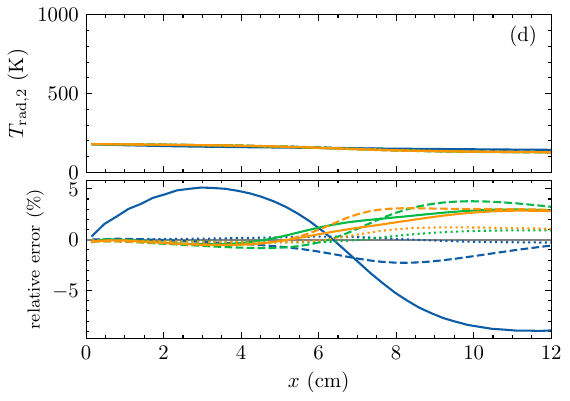}\\
    \includegraphics[width=0.5\textwidth]{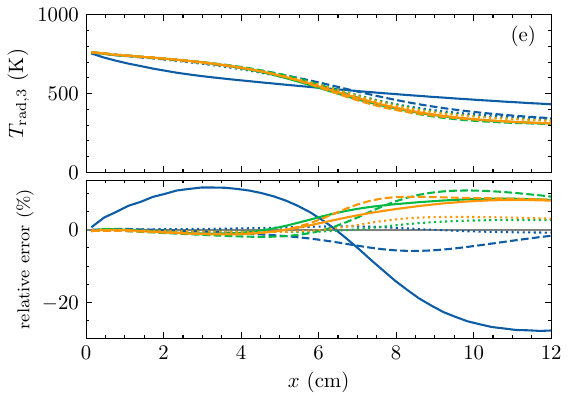}\includegraphics[width=0.5\textwidth]{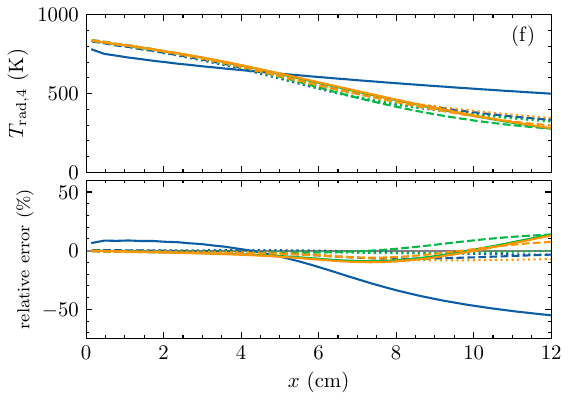}
    \caption{Results for the Marshak wave tests with a continuously-variable opacity $\chi_0 \propto \nu^{-2}$ obtained from calculations with 4, 8, and 16 energy bins using various opacity models -- PC (blue), PPL with fixed slopes (green), and PPL with free slopes (orange). Solid, dashed, and dotted lines show results for 4, 8, and 16 radiation groups, respectively. The top row shows the gas temperature and the total radiation temperature, while the bottom row shows the radiation temperature of the four radiation groups, $[6\times 10^{10}, 6\times 10^{11}, 6\times 10^{12}, 6\times 10^{13}, 6\times 10^{14}]$ Hz. In each panel the error shown is relative to the converged results we obtain at very high spectral resolution.}\label{fig:marshak2}
\end{figure*}

Our next test is intended to measure the accuracy and convergence rates of the three different methods we have described for computing group-mean opacities: PC, PPL with fixed slopes for the radiation quantities, and PPL will full-spectrum fitting for the radiation quantities using the algorithm outlined in \aref{sec:slope}. In order to test these approaches we repeat the Marshak wave test, but instead of the piecewise constant opacities adopted in \citet{Vaytet2011} we use a more realistic opacity that varies continuously as function of frequency, $\chi_0(\nu) = 3.2 \times 10^{4} (\nu / 10^{13} ~{\rm Hz})^{-2}~{\rm cm}^{-1}$. For each opacity method, we run the same simulation with 4, 8, 16 energy bins, and compare their results with the converged result, which we obtain by using 128 bins with the PC method -- though at this frequency resolution all three methods converge to nearly-identical results. For these tests frequency bins are evenly distributed in logarithmic space between $6 \times 10^{10}$ Hz and $6 \times 10^{14}$ Hz. 

We show the results in \autoref{fig:marshak2}. Panels (a) and (b) show $T_{\mathrm{rad},g}$ and $T_{\rm rad}$, respectively, while the remaining panels (c) - (f) show effective radiation temperature, $T_{\rm g}$ (defined per \autoref{eq:Tgdef}), integrated over four energy ranges: $[6\times 10^{10}, 6\times 10^{11}, 6\times 10^{12}, 6\times 10^{13}, 6\times 10^{14}]$ Hz. These bins correspond to the ones used in the four-group simulation, and in simulations with more than four bins we simply sum over the finer bins to allow direct comparison. 

From panel (a) we immediately notice a few trends on the convergence rate of these three methods. First, with 4 bins, the PPL-fixed and PPL-free methods work similarly and significantly better than the PC method. With 8 bins, all three methods give results with similar accuracy. With 16 bins, the results from the PPL-fixed and PPL-free methods are similar and slightly worse than that of the PC method. These trends are understandable. With a small number of bins, the resolution of the spectrum is poor, so the PPL-fixed method, which uses the average slopes for the spectrum, works as well as or better than the PPL-free method, and both of them are better than the PC method. However, when the spectrum is well resolved with a large number of bins, while the PPL-free methods provides a good approximation to $\chi_{0B,g}$ and $\chi_{0E,g}$, the diffusion flux-mean opacity (\autoref{eq:chiDiff}) introduces inaccuracies into the calculation. Our tests show that this inaccuracy is diminished when using PPL models for evaluating $\chi_{0F,g}$. Conversely, because the bin width is so small, the piecewise-constant opacity model is a good approximation so the PC method work the best. Finally, we note that the overall accuracy is extremely good for even a small number of bins: for example four bins using the PPL-fixed method is sufficient to achieve accuracy better than 10\% in all quantities.

Based on this test, we recommend the following strategy for choosing the opacity method based on the problem and the number of energy groups. For $N_g \lesssim 10$ we recommend using PPL with fixed-slope spectrum as it has the best accuracy. For $N_g \gtrsim 10$ we recommend PC, provided that the bin widths are small enough, on the basis that it is nearly as accurate as PPL-free, but is noticeably faster since there is no need to evaluate the slopes. More precisely, we recommend PC whenever the bin widths in logarithmic space are smaller than $\sim W / 10$, where $W$ is the width of the full frequency range in logarithmic space, and PPL with fixed-slope otherwise.

\subsubsection{Linear diffusion with multigroup radiation}\label{sec:linear}

\begin{figure*}
    \centering
    \includegraphics[scale=0.8]{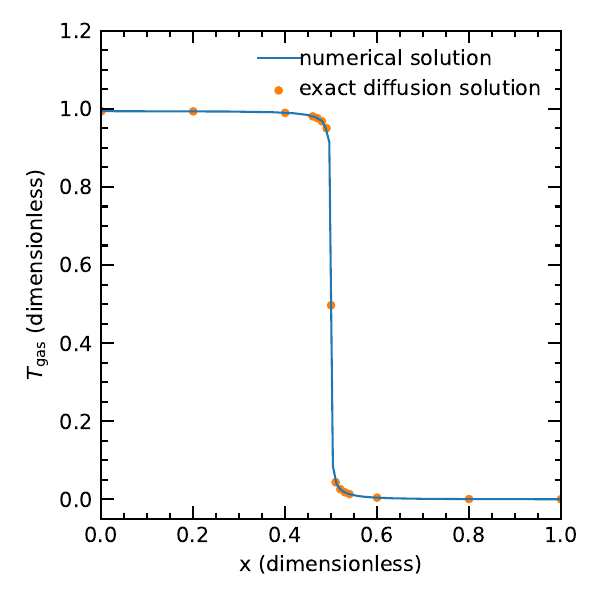}\hspace{-0.2cm}\includegraphics[scale=0.8]{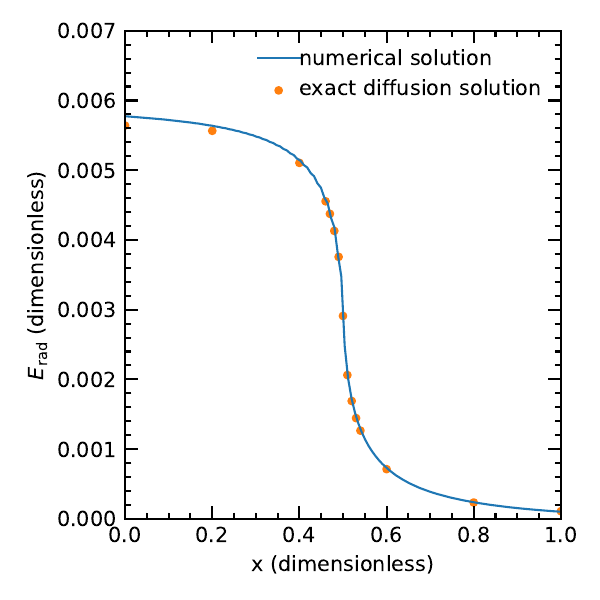}
    \caption{Linear multigroup diffusion test discussed in \autoref{sec:linear}. We show dimensionless temperature (left) and total radiation energy density (right) at the dimensionless time $t = 1$. \newnew{The optical depth per cell is $\tau_c \approx 6\times 10^7$ for the lowest frequency group.} The numerical results (solid lines) are in good agreement with the exact {\it diffusion} solutions (circles). We observe a slight mismatch in the radiation energy density between our numerical solution and the exact diffusion solution because the problem is not in the diffusion limit for highest frequency groups -- see main text for details. }
    \label{fig:linear}
\end{figure*}

Our third pure radiation test is a linear diffusion problem introduced by \citet[][their problem 1]{Shestakov2005}. This test is unique in that it includes non-trivial frequency dependence in the opacity and radiation field structure, but also admits an exact analytic solution. An additional benefit specific to our code is that this test is derived in the diffusion limit, and therefore allows us to verify that our multigroup method shares the asymptotic-preserving property demonstrated for the single-group formulation in \citetalias{He2024}, meaning that, although we are using a two-moment method that can properly capture both the streaming and diffusion limits, our scheme properly approaches the diffusion limit when the optical depth is high.

For this test we use non-dimensional units where $c = \sqrt{3}$, $8\pi h / c^3 = 1$, $k_B = h$, and $a_R = 1$. The computational domain is $x=[0, 4]$, with a reflecting boundary condition at the lower boundary and an outflow boundary condition at the upper boundary; we discretise this domain into 256 equal-sized cells. This domain is filled with a motionless gas with density $\rho = 1$, heat capacity $C_V = 1$, and temperature $T = 1$ for $x < 0.5$ and $T=0$ for $x > 0.5$; the initial radiation energy density is zero everywhere.

We use the same group structure as \cite{Shestakov2005}: 64 groups with $\nu_{1^-} = 0$ and $\nu_{1^+} = 5 \times 10^{-4}$, and the frequency width of all other groups set by $\nu_{g^+} - \nu_{g^-} = 1.1 [\nu_{(g-1)^+} - \nu_{(g-1)^-}]$, i.e., the width of group $g$ is set to be 1.1 times the width of group $g-1$.  The absorption coefficient is taken to be constant within each group, and is set to a value $\chi_{0,g} = \nu_g^{-3} / \sqrt{3}$ where $\nu_g \equiv \sqrt{\nu_{g-1 / 2} \nu_{g+1 / 2}}$ for all groups except $g=1$, for which we adopt $\nu_1 = 0.5 \nu_{1^+}$. For our chosen resolution, this means that the optical depth per cell is $\tau = 6\times 10^7$ at $\chi_0 = \chi_{0,1}$, and $\tau = 2 \times 10^{-4}$ at $\chi_0 = \chi_{0,64}$. Therefore at most frequencies the problem is strongly in the asymptotic diffusion limit where the photon mean free path is unresolved; however, at the highest frequencies the optical depth per cell becomes small, and in fact the optical depth across the entire domain falls to $\approx 0.25$. In order to obtain a closed-form solution, \citet{Shestakov2005} assume purely diffusive radiation transport, and consider a medium for which radiation emission follows a modified Wien's law rather than Planck's law. For group $g$, the emissivity is 
\begin{equation}\label{eq:Bmod}
    B_g=\frac{2 \pi k_B}{c^2} \nu_g^3\left[\exp \left(-\frac{h \nu_{g-1/2}}{k_B T_f}\right)-\exp \left(-\frac{h \nu_{g+1 / 2}}{k_B T_f}\right)\right] T,
\end{equation}
where $T_f=0.1$, and for the purposes of this test we modify \textsc{quokka} to use this emission function.

We simulate the system using a CFL number of 0.8 and run to time $t=1$. We show the numerical results in \autoref{fig:linear} and compare them with the tabular data of \cite{Shestakov2008}. We plot the domain between 0 and 1 to centre the location of discontinuity. The results for gas temperature are in excellent agreement with the exact diffusion solution, confirming that our multigroup method retains the asymptotic-preserving property demonstrated for our the frequency integrated method in \citetalias{He2024}, i.e., we recover the diffusion limit even when we do not resolve the photon mean free path. 
We see a slight mismatch in the radiation energy density at $x \lesssim 0.2$ between our result and the reference solution, but this is to be expected. Since the optical depth across the domain is $<1$ in the highest frequency bins, this is not a pure diffusion problem, contrary to the assumption used to obtain the closed-form solution. Our two-moment code does not assume pure diffusion, unlike previous authors who have used this test \citep[e.g.,][]{Shestakov2008, Zhang2013}, so the difference is very likely due to the more accurate treatment of radiation transport that we perform. This hypothesis is consistent with where the mismatch between the analytic diffusion and numerical solutions occurs: the diffusion approximation is best in the low-temperature region, where most radiation energy is at lower frequencies where the opacity is highest, and worst in the high-temperature region, which is precisely where our numerical solution differs from the exact diffusion one. It is also consistent with the results of the Su-Olson test presented in \citetalias{Wibking2022} (their Figure 10), where the $M_1$ solution obtained by \quokka{} yields a slightly higher radiation energy density than a diffusion approximation solution, while more closely matching the transport solution in the high-temperature region.

\subsection{Radiation-hydrodynamic tests}
\label{ssec:rhd_tests}

Our tests thus far have been problems of pure radiative transfer with a time-independent gas background. We now expand our testing regime to full RHD.

\subsubsection{Frequency-dependence of the Doppler effect}\label{sec:doppler} 

\begin{figure*}
    \centering
     \includegraphics[width=0.48\textwidth]{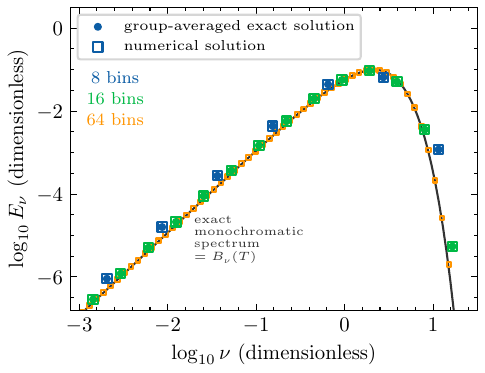}    \includegraphics[width=0.48\textwidth]{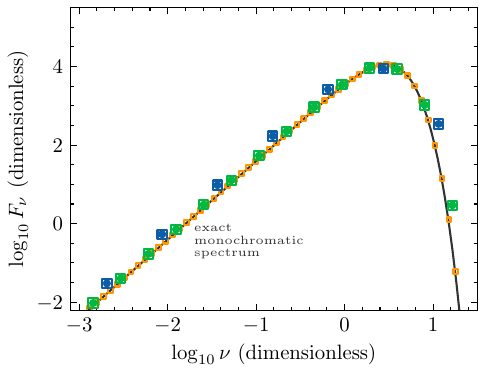}
    \caption{Solutions for the monochromatic radiation energy density (left) and flux (right) in the lab frame from the problem described in \autoref{sec:doppler}. 
    Lines show the exact monochromatic solution, while filled points show the mean energy (left) and flux (right) in each group computed by integrating the exact solution over the same frequency bins used in the simulation. Open squares show the corresponding group-mean energy and flux computed by the simulation, defined as
    $E_{\nu,g} = E_g / (\nuplus - \numinus)$ and similarly for the flux. Blue points show results for 8 frequency bins, green for 16 frequency bins, and orange for 64. We see that the group-averaged numerical solutions (open squares) show excellent agreement with the group-averaged exact solutions (points) regardless of the number of energy bins; note that we refrain from plotting the exact frequency-integrated solutions (filled points) for $N_g = 64$ because they are visually indistinguishable from the monochromatic exact solution line, and it is therefore simpler to compare out $N_g = 64$ solution directly to the exact monochromatic one.
}
    \label{fig:doppler}
\end{figure*}

Variations in the spectrum of radiation produced by moving matter due to the Doppler effect are handled by the $\partial/\partial \nu$ terms in \autoref{eq:cG0} and \autoref{eq:cGi}, and these terms are at the heart of our mixed-frame formulation, since they avoid the need to carry out explicit frame transformations of the radiation quantities. In this test, we check that the Lorentz transformation of emissivities and opacities is correctly implemented in our numerical scheme by considering a moving medium in thermal equilibrium and verifying that we recover the correct, Doppler-shifted spectrum.

For this test we adopt a dimensionless unit system for which $a_R = k_B = h = c = 1$. Our initial condition consists of a uniform medium with density $\rho = 1$, temperature $T_\mathrm{gas,0} = 1$, specific heat at constant volume $C_V = 3/2$, velocity $v = 0.001$, and frequency-independent opacity $\chi_0 = 10^5$. The initial radiation energy density is zero everywhere. The gas occupies a 1D periodic domain covering the region $[0, 64]$, which we cover with 64 grid cells. We use either 8, 16, or 64 frequency groups evenly covering $\nu = [10^{-3}, 10^2]$ logarithmically; for a frequency-independent opacity, the PC and PPL methods are identical, so it does not matter which of the approaches outlined above we choose. We advance the system to time $t = 1000$, long enough to reach steady-state given the very high opacity, using a hydro CFL number of 0.8 and a radiation CFL number of 8.0; the latter is permissible because our code is asymptotic-preserving \citepalias{He2024}. 

We can compute the expected equilibrium solution as follows. First note that, in the frame comoving with the gas, the gas and radiation should reach equal temperatures, and the equilibrium temperature can be calculated directly from energy conservation once we recall that gas temperature is a Lorentz scalar, and thus is the same in all frames: 
\begin{equation}
    a_R T^4 + C_V \rho T = C_V \rho T_{\rm gas,0},
\end{equation}
where $T_{\rm gas,0} = 1$ is the initial gas temperature. The resulting equilibrium temperature is $T = 0.768032502191$. The difference between radiation energy density in the lab and comoving frames is second-order in $v/c$, so for our first-order treatment the equilibrium radiation spectrum $E_\nu$ in both the lab and comoving frames should simply be a Planck function $B_\nu(T)$ evaluated at the equilibrium temperature. We verify that our code reproduces this solution in the left panel of \autoref{fig:doppler}, which shows near-perfect agreement independent of the number of frequency bins. The equilibrium gas temperature we obtain is 0.7680327097575, which differs by $3 \times 10^{-7}$ from the exact expectation. 

While this result is encouraging, the spectrum of the radiation flux is a far more stringent test. In the frame comoving with the matter, the radiation is isotropic and the flux is zero. In the lab frame, the intensities are Doppler-shifted depending on the angle, causing a combination of Doppler and aberration effects that skew specific intensities in the direction of gas motion, resulting in non-zero radiation flux. We can calculate the expected flux spectrum by integrating the Doppler-shifted specific intensity over angle. To order $\beta = v/c$, the transformation of frequency from the lab frame to the fluid frame is given by \citep{Mihalas2001} 
\begin{equation}
    \nu_0 =\nu\left(1-\beta^i n^i\right)
\end{equation}
and the specific intensity transformation from the fluid frame to the lab frame is given by 
\begin{equation}
    I(\vecn, \nu) = \frac{\nu^3}{\nu_0^3} I_0\left(\vecn_0, \nu_0\right).
\end{equation}
The radiation flux is therefore
\begin{equation}\label{eq:bbF}
\begin{split}
F_\nu^i & = \int d \Omega \ n^i I(\vecn, \nu) \\
& = \int d \Omega \ n^i \frac{\nu^3}{\nu_0^3} I_0(\vecn_0, \nu_0) \\
& = \int d \Omega \ n^i \frac{\nu^3}{\nu_0^3} B(\nu_0) \\
& = \int d \Omega \ n^i \frac{2\nu^3}{e^{(\nu / T) \left(1-\beta^i n^i\right)}-1} \\
& = \int_0^{\pi} d \theta \int_0^{2 \pi} d \varphi \sin^2 \theta \cos \varphi 
\frac{2\nu^3}{e^{(\nu / T)(1-\beta \sin \theta \cos \varphi)}-1}
\end{split}
\end{equation}
where we have used $\beta^i = (\beta, 0, 0)$. The integral \autoref{eq:bbF} can be evaluated numerically.\footnote{One can alternatively derive the group-averaged fluxes directly from \autoref{eq:Fdiff}; we choose to evaluate it as shown above to better illustrate the physical origin of the effect. However, one can readily verify that the result obtained from numerical evaluation of this integral is identcal that derived from \autoref{eq:Fdiff} to at least order $v/c$, as expected.} We show this comparison in the right panel of \autoref{fig:doppler}, which again shows excellent agreement between the analytical calculation and the numerical results. The maximum relative error between the numerical solution and the exact solution is below $0.1\%$, regardless of the number of energy bins. 
The excellent agreement between our numerical simulation and the analytical results on the spectra demonstrates that our multigroup implementation accurately accounts for frequency-dependent Doppler shifts to order $v/c$.

\subsubsection{Multigroup non-equilibrium radiation shock} 

\begin{figure*}
    \centering
    \includegraphics[width=0.8\columnwidth]{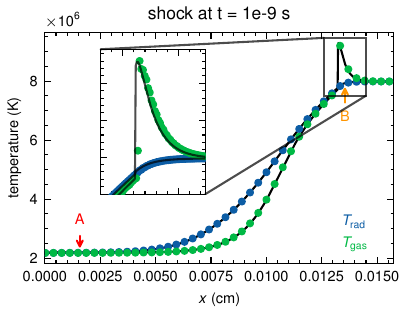}
    \includegraphics[width=1.2\columnwidth]{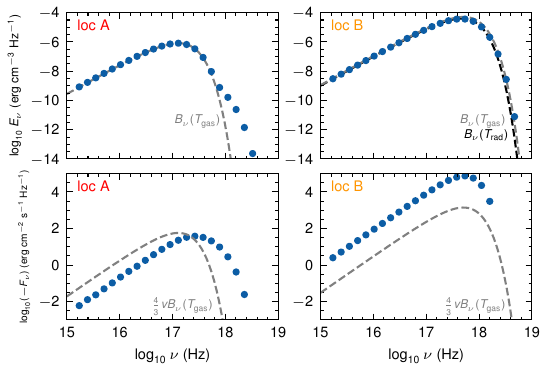}
    \caption{Multigroup non-equilibrium radiation shock with $\mathcal{M} = 3$. {\it Left}: Temperature profile at $t=10^{-9}$ s. We compare radiation and gas temperature from the numerical calculation (dots) with the exact steady-state solution (solid lines), finding excellent agreement. {\it Right}: Spectra of the radiation energy density and radiation flux at the two marked locations in the profile. Dots show the numerical solution, and for comparison dashed lines in the top and bottom panels show $B_\nu(T_\mathrm{gas})$ and $(4/3) v B_\nu(T_\mathrm{gas})$, respecitvely, where $v$ and $T_\mathrm{gas}$ are the gas velocity and temperature at the indicated location.}
    \label{fig:shock} 
\end{figure*}

Our next RHD test is the classical non-equilibrium radiative shock test, following the setup used by \cite{Skinner2019} for the Mach ${\cal M} = 3$ test given by \cite{Lowrie2008}. In \citetalias{He2024}, we showed results computed with the grey RHD solver. Here, we present results from a simulation with the multigroup RHD solver along with the resolved radiation spectrum, demonstrating both that our method properly reproduces the frequency-integrated result when the opacity is frequency-independent, and that it returns a realistic radiation spectrum while doing so.

The setup of this test is exactly the same as that in \citetalias{Wibking2022} except that we now add a new run with 32 radiation groups. The lowest and uppermost boundaries of the groups are at $10^{15}$ and $10^{20}$ Hz, respectively, and the groups are evenly spaced in logarithm. We use an adiabatic equation of state with an adiabatic index to $\gamma = 5/3$. The shock is simulated in a 1D region with $x \in [0, x_R]$, where $x_R = 0.01575$ cm, resolved with 512 cells, with the discontinuity placed at $x_0 = 0.0130$ cm. The gas conditions on the left and right sides of the shock are uniform, with densities, temperatures, and velocities given by $\rho_L = 5.69$  g cm$^{-3}$, $T_L = 2.18 \times 10^6$ K, $v_L = 5.19 \times 10^7$  cm s$^{-1}$, and  $\rho_R = 17.1$ g cm$^{-3}$, $T_R = 7.98 \times 10^6$  K, $v_R = 1.73 \times 10^7$ cm s$^{-1}$, respectively. We initialise the radiation energy densities in each group to the values appropriate for blackbody radiation at a temperature equal to the gas temperature, and the radiation fluxes to zero. The states at spatial boundaries are held at fixed values. Following \cite{Skinner2019}, we use a reduced speed of light $\hat{c} = 10 (v_L + c_{s,L})$, where $c_{s, L}$ is the adiabatic sound speed of the left-side state. The opacity of the gas is $\chi_0 = 577 \ {\rm cm^{-1}}$ independent of frequency (so that the choice of opacity model does not matter) and local gas properties, and the mean molecular weight is $\mu = m_{\rm H}$. To match the assumptions used in the semi-analytic solution, we use the Eddington approximation, $\tenP = (1/3) E \tenI$, to calculate the radiation pressure tensor. We use a CFL number of 0.4 and evolve until $t=10^{-9}$ s. 

In the left panel of \autoref{fig:shock} we show the temperature profile at the end of simulation, with an inset zooming in on the Zel'Dovich spike near the shock interface. We compare our numerical solution with the semi-analytic, exact solutions of \cite{Lowrie2008}, finding excellent agreement in both the non-equilibrium spike region and in the radiatively heated shock precursor in the upstream area. The $L_1$ norm of the relative error of the gas temperature is 0.38 per cent, which is as good as the multigroup solution of \cite{Skinner2019}.  This demonstrates that we successfully reproduce the frequency-integrated result.

With our multigroup method, however, we can go further by resolving the spectrum of the radiation energy density and flux. We plot spectra at two positions marked as A and B in \autoref{fig:shock}. Location A is far upstream of the shock at $x/x_R = 0.1$, and location B is just right of the shock front. We see that the radiation deviates from a blackbody spectrum at both locations. At location A, the energy spectrum is close to a blackbody at the local gas temperature for frequencies at and below the peak, but there is an excess at high frequency. This excess is due to radiation emitted by the hot material downstream of the shock, which propagates upstream to location A, as indicated by the negative value of the radiation flux at all frequencies. The optical depth from the shock to location A is moderate, $\tau\approx 7$, so enough of the flux reaches A to be noticeable in the energy spectrum at frequencies above the local blackbody peak, where emission from local material is exponentially suppressed. By contrast, location B is near the temperature maximum across the entire domain, so there is no significant source of radiation from other locations, and the radiation energy density is close to an unperturbed blackbody curve. By contrast there is a very large negative (left-propagating) radiation flux due to the difference in radiation temperatures in the upstream and downstream regions.

Overall, the multigroup non-equilibrium radiation shock test validates the robustness and accuracy of our multigroup RHD solver, demonstrating excellent agreement with semi-analytic solutions and effective resolution of the radiation spectrum across different regions of the shock profile.

\subsubsection{Advecting radiation pulse with variable opacity}
\label{sec:pulse}

\begin{figure*}
    \centering
    \includegraphics[width=0.9\textwidth]{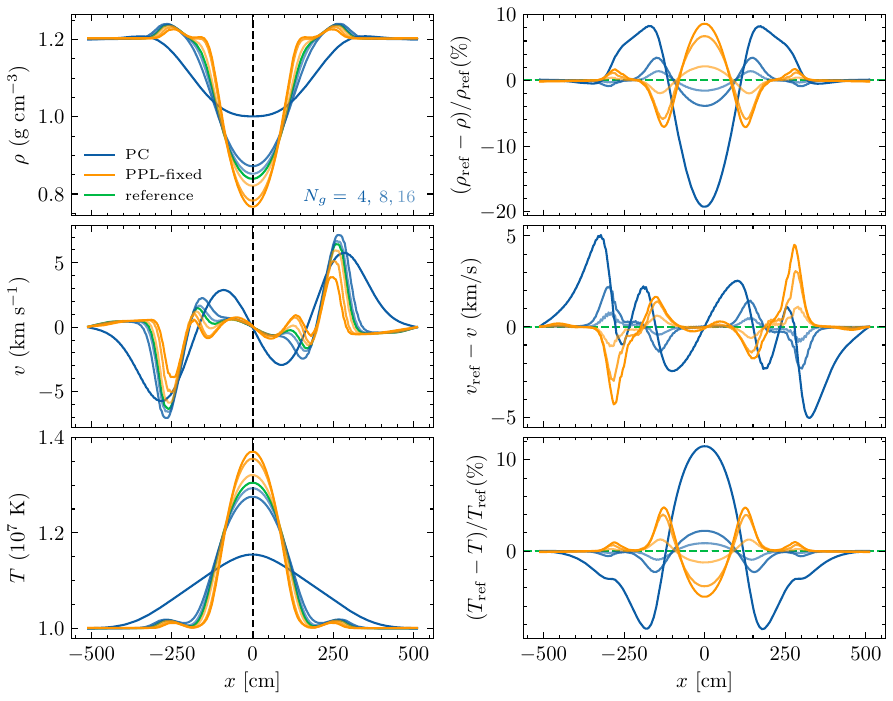}
    \caption{Results from the multigroup diffusing, advecting radiation pulse problem with frequency- and temperature-dependent opacity using two opacity models: piecewise constant (PC, blue), and PPL with fix-slope spectrum (PPL-fixed, green). \newnew{This problem is in the asymptotic diffusion limit -- the optical depth per cell is $\tau_c \approx 200$.} In the left column, green solid lines show the reference solution computed using the analytically-evaluated frequency-averaged opacities, while the three blue and orange lines in each panel show results computed using the PC and PPL-fixed methods with 4, 8, or 16 energy groups; results with more groups are lighter and lie closer to the reference solution. The black dashed vertical line shows $x=0$, and is provided as a visual aide to help judge the degree of symmetry between the leading and trailing edges of the pulse in the advecting case. In the right column, blue and green lines show errors in the multigroup results relative to the reference solution, with the black dashed horizontal line at zero indicating perfect agreement. We see that for $N_g = 4$ the results of the PPL-fixed method are noticeably better than those from the PC method, but this advantage disappears when a large number ($\gtrsim 8$) of radiation groups are used. We also test both models with 64 energy groups, and find that both perfectly match the reference solution; we omit this curve from the plot, since it is visually identical to the reference solution. This verifies that our method correctly approaches the exact Planck- and Rosseland-mean opacities at high optical depth.
    } 
    \label{fig:pulsem1}
\end{figure*}

In \citetalias{He2024}, we performed a test of advecting a radiation pulse by the motion of matter \citep[following the original test problem introduced by][]{Krumholz2007} with the grey RHD solver. Here we repeat the test with frequency- and temperature-dependent opacities following the extension to the test introduced by \citet{Zhang2013}. This test has two main purposes: first, it allows us
to verify that our method correctly reduces to the diffusion limit, and that our approximate opacities correctly approach the Planck- and Rosseland-means, when the optical depth is large. Second, it allows us to compare the performance of our piecewise powerlaw approximation to the traditional piecewise constant approximation in a case where the true functional form of the frequency dependence is not a pure powerlaw, but instead has significant curvature in log-log space.

As in \citetalias{He2024}, we consider a 1D medium with initial temperature and density profiles 
\begin{gather}\label{eq:pulseT}
T=T_0+\left(T_1-T_0\right) \exp \left(-\frac{x^2}{2 w^2}\right), \\
\rho=\rho_0 \frac{T_0}{T}+\frac{a_R \mu}{3 k_{\mathrm{B}}}\left(\frac{T_0^4}{T}-T^3\right),
\end{gather}
with $T_0=10^7 \mathrm{~K}, T_1=2 \times 10^7 \mathrm{~K}, \rho_0=1.2 \mathrm{~g} \mathrm{~cm}^{-3}, w=24 \mathrm{~cm}$, and $\mu=2.33 m_p=3.9 \times 10^{-24} \mathrm{~g}$. For this choice of parameters the system is initially in pressure balance, with radiation pressure dominating in the high-temperature region near $x=0$ and gas pressure dominating elsewhere. Thus without radiation transport the system should remain static. However, due to radiation diffusion, pressure balance is lost and the gas moves. We run two versions of the simulation: a non-advecting case where the initial velocity is $v_0 = 0$ everywhere, and an advecting case where $v_0 = 10^6 \ {\rm cm~s^{-1}}$ everywhere. Following \citet{Zhang2013}, who developed a multigroup version of this test, the matter opacity is a continuous function of temperature and frequency, given by \begin{equation}\label{eq:pulsek}
    \chi_0 = 180\left(\frac{T}{10^7 \mathrm{~K}}\right)^{-0.5}\left(\frac{\nu}{10^{18} \mathrm{~Hz}}\right)^{-3}\left[1-\exp \left(-\frac{h \nu}{k_B T}\right)\right] \mathrm{cm}^{-1} .
\end{equation}
The reason for adopting this particular functional form will become apparent below. The optical depth across the pulse at $T = T_0$ for frequences near the peak of the Planck function is $\tau \sim 10^3$, and since $\beta \equiv v_0/c = 3.3 \times 10^{-5}$, we have $\beta \tau \sim 0.1$, placing this problem in the static diffusion limit for both the advecting and non-avecting cases.

In the initial condition for the non-advecting case, radiation and matter are in thermal equilibrium and radiation flux is zero; we therefore set our initial values of $E_g$ by numerically integrating the Planck function at the local matter temperature in each cell over the frequency range of each group, and set our initial values of $F_g$ to zero. In the advecting case, as discussed in \autoref{sec:doppler}, the radiation energy density remains equal to the Planck function (to order $v/c$), but due to the Doppler effect the radiation flux is non-zero and can be computed using \autoref{eq:Fdiff}. We therefore initialize $E_g$ as in the non-advecting case, and $F_g$ from \autoref{eq:Fdiff}.

The computational domain for this test is a 1D region spanning $(-512, 512)$ cm with periodic boundaries, and the grid consists of 256 cells. \newnew{The optical depth per cell for frequences near the peak of the Planck function is $\tau_c \approx 200$, placing this problem in the asymptotic diffusion limit.} For both the advecting and non-advecting case, we perform two sets of multigroup simulations using the PC and PPL-fixed models; we omit PPL-free based on our recommendations from \autoref{sec:marshak2}. Each set consists of three runs with 4, 8, and 16 radiation groups. The radiation bins are distributed evenly in logarithmic space between $10^{15}$ Hz and $10^{19}$ Hz. We do not use the Eddington approximation, and instead compute the radiation pressure for each group using the M1 closure.

We can obtain a reference solution to which to compare the simulation results as follows. First note that, for the functional form of the opacity $\chi_0$ that we have chosen, we can evaluate the frequency-integrated Planck- and Rosseland-mean opacities analytically:
\begin{equation}
    \chi_{(0P,0R)} = (3063.96, 101.248) \left(\frac{T}{10^7 \mathrm{~K}}\right)^{-3.5} \mathrm{~cm}^{-1}.
\end{equation}
Because these mean opacities are exact, we can use the grey RHD method described in \citetalias{He2024} to compute a reference solution representing the results toward which our finite frequency-resolution calculations should converge as $N_g \to \infty$. We therefore compute a reference solution using the grey method described in \citetalias{He2024} and the exact analytic expressions for $\chi_{0P}$ and $\chi_{0R}$; for consistency with our order $v/c$ method here, we use the order $v/c$-accurate grey method as well (see \citetalias{He2024} for details). We compute reference solutions for both the advecting and non-advecting cases, though as shown in \citetalias{He2024} the results for these are nearly identical once we remove the overall translation. Our reference solution agrees well with the \newnew{diffusion solution} obtained by \citet{Zhang2013}.

We show the density, temperature, and velocity profiles from the advecting runs at $t = 2 w / v_0 = 4.8 \times 10^{-5}$ s in \autoref{fig:pulsem1}; the results are shifted in space by $v_0 t$ to center them. We refrain from plotting the corresponding results from the non-advecting runs because the differences are so small as to be indistinguishable to the eye, demonstrating the accuracy of our multigroup scheme in capturing radiation advection. Note that our advecting solutions are also almost perfectly symmetric about the $x=0$ line, demonstrating that, as with the grey method from \citetalias{He2024}, our method here captures advection effects without introducing any artificial asymmetry between the leading and trailing edges of the pulse. 

Comparing our multigroup solutions to the reference solution, we see that our results are overall very accurate when the frequency resolution is good, regardless of whether we use PC or PPL, \newnew{demonstrating that our multigroup scheme accurately preserves the asymptotic diffusion limit, similar to the single-group scheme in \citetalias{He2024}.} For $N_g = 16$, we match the reference solution to typical accuracies of a few percent for $\rho$ and $T$, and within 1 km s$^{-1}$ for $v$ (where percentage errors are not easily defined, since the reference solution has $v=0$ at several points). However, for $N_g = 4$ the results of the PPL-fixed model are noticeably better than those obtained using the PC method, with errors $2-3$ times smaller; errors for PPL with $N_g = 4$ are comparable to errors for PC with $N_g = 8$. This demonstrates that with PPL we have achieved our goal of substantially improving accuracy for simulations with modest frequency resolution without increasing the computational cost significantly. When we use more frequency bins, the accuracy of both methods are similar. This reinforces our recommendation from \autoref{sec:marshak2} that one should use PC for $N_g \gtrsim 10$, since in this case PPL offers no accuracy gain and thus one might as well use the (slightly) computationally-cheaper PC method, but that for fewer bins PPL-fixed is preferable due to its higher accuracy.

\subsubsection{Radiation sphere in static equilibrium}

\begin{figure}
    \centering
    \includegraphics[width=\columnwidth]{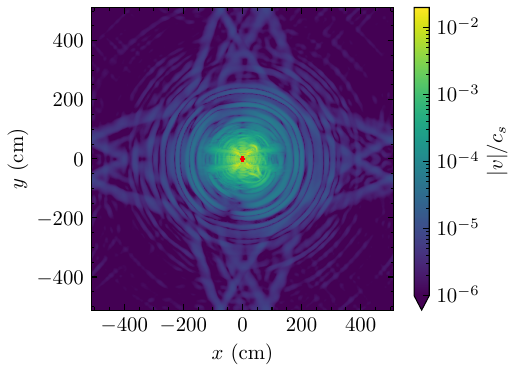}
    \caption{Magnitude of velocity at $t = 4.8 \times 10^{-5}$ in the multigroup static equilibrium test after subtracting off the initial background advection velocity. The exact solution should have $v=0$ everywhere. The sound speed in this problem is 267 km s$^{-1}$ and the advection speed is 10 km s$^{-1}$. The maximum velocity we find is about 1 per cent of the sound speed, indicating good preservation of the static equilibrium in multiple dimensions.}
    \label{fig:static}
\end{figure}

In our final test, we demonstrate that the multigroup RHD solver can maintain a stable static equilibrium in multiple dimensions and in a situation where both radiation pressure and gas pressure are dominant over differnent parts of the domain. The setup is the same as the static equilibrium simulation of \cite{Zhang2013}. The initial conditions for radiation and matter are the same as those in the test of the advecting radiation pulse test (\autoref{sec:pulse}) except that the coordinate $x$ in \autoref{eq:pulseT} is replaced by $r = \sqrt{x^2 + y^2}$, so that the region dominated by radiation pressure becomes circular, and the opacity increased to $\kappa_0 = 10^{20} \ {\rm cm^2~g^{-1}}$. We use a 2D grid of $-512 \ {\rm cm} < x < 512 \ {\rm cm}$ and $-512 \ {\rm cm} < y < 512 \ {\rm cm}$ with 512 uniform cells in each direction. The initial velocity is $v_x = 10^6 \ {\rm cm~s^{-1}}$ and $v_y = 0$ everywhere. The very large specific opacity ensures that negligible radiation diffusion occurs over the course of the simulation, so the radiation and matter should remain in pressure balance and the only velocity should be due to the initial advection. 

We use 4 radiation groups evenly spaced in logarithmic space from $10^{16}$ to $10^{20}$ Hz. \autoref{fig:static} shows the magnitudes of velocity at $t = 4.8 \times 10^{-5}$ s, enough time for the pulse to have been advected twice across its initial width. The maximum velocity at this time, after subtracting off the initial advection velocity $v_x$, is about $1 \%$ of the typical sound speed, or $3 \times 10^5 \ {\rm cm~s^{-1}}$. Such a relatively small gas velocity indicates that the multigroup solver in \quokka{} can maintain a good static equilibrium in multiple dimensions, even though the radiation pressure and the gas pressure are operator-split in the Riemann solver. 

\section{Conclusion}\label{sec:conclusion}

We have presented an extension of the \quokka{} code to incorporate multigroup RHD in a mixed-frame formulation. Our approach successfully integrates the advantages of lab-frame radiation transport with comoving-frame emissivities and opacities, ensuring exact conservation of energy and momentum and addressing the complexities of frequency-dependent radiation-matter interactions. To our knowledge this work represents the first mixed-frame, moment-based, multigroup method presented in the astrophysical literature. 

The equations we derive to describe the radiation four-force in the multigroup method, \autoref{eq:G0_3} and \autoref{eq:G1_3}, are relatively simple and can be expressed in terms of group-integrated radiation quantities, group-mean opacities, and the opacity at group boundaries. This offers the significant advantage that in our method we can treat matter-radiation coupling purely locally, in a way that requires no non-local implicit steps. As a result, the entire code maintains the same communication requirements as a pure hydrodynamics update, making it highly efficient for parallel computations on GPU architectures. The source term is handled with a set of equations where the Jacobian matrix is of size $N_g + 1$, where $N_g$ is the number of radiation groups. The inversion of this matrix requires only $O(N_g)$ complexity, ensuring that the overall computational complexity of the radiation solver scales linearly with the number of radiation groups. 

A second key innovation of our method is the novel piecewise power-law approximation, 
which we introduce for the purpose of calculating the various group-averaged opacities that appear in the radiation-matter exchange terms. Construction of this scheme requires some care to ensure that it retains the correct limiting behaviour at high optical depths, but once we satisfy this constraint, we find that the new scheme offers significantly better accuracy at only marginally greater cost than the traditional approach of approximating the opacity as constant within a frequency bin when the number of frequency groups is $\lesssim 10$. Such coarse frequency resolution is often unavoidable due to computational constraints, and thus the new scheme is often preferable in practice.

Through a series of rigorous tests, we demonstrate that our multigroup method maintains the asymptotic-perserving properties of the original \newnew{single-group} scheme \newnew{in the diffusion limit}, accurately recovers all relevant limits of RHD, effectively handles variable opacity using our novel piecewise powerlaw approximation, and enables spectrum-resolving capability. We also highlight the superiority of the piecewise powerlaw method over the traditional piecewise-constant approach.

Future work will focus on extending the capabilities of \quokka{} to include additional physical processes, such as photoionisation, and further optimizing the algorithm for large-scale parallel computations on GPU architectures. The development of new methods for handling scattering and non-local thermodynamic equilibrium conditions will also be explored.

\section*{Acknowledgements}

CCH and MRK acknowledge support from the Australian Research Council through Laureate Fellowship FL220100020. This research was undertaken with the assistance of resources and services from the National Computational Infrastructure (NCI) and the Pawsey Supercomputing Centre, which are supported by the Australian Government, through award jh2. The \quokka{} code is based on the AMReX module \citep{AMReX_JOSS}. The analysis made signiﬁcant use of the package matplotlib \citep{Hunter:2007}. 

\section*{Data Availability}
 
The \quokka{} code, including the source code and summary result plots for all tests presented in this paper, is available from \url{https://github.com/quokka-astro/quokka} under an open-source license.

\bibliographystyle{mnras}
\bibliography{IMEX,refs} %

\appendix

\section{On the evaluation of powerlaw indices of radiation quantities}
\label{sec:slope}

\begin{figure}
    \centering
    \includegraphics[]{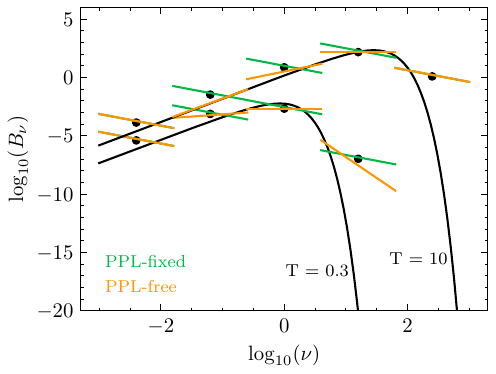}
    \caption{Demonstration of reconstructing powerlaw indices of radiation quantities. The black solid curves are Planck functions at the indicated temperatures, representing the real spectra. The dots are the group-integrated quantities $E_g$ using 5 radiation groups, which are the state variables used in the simulation. All quantities are dimensionless. The green and orange lines represent the reconstructed spectra from the state variables using the PPL method with fixed slope and with full-spectrum fitting, respectively. The green and orange lines overlap in the first and last group.}
    \label{fig:slopes}
\end{figure}

In our PPL with full-spectrum reconstruction method, we must with the powerlaw slopes of $\alpha_{Q_g}$ of all radiation quantities on fly from the time-evolving and position-dependent solution. We do so as follows. We first calculate the slopes at the bin edges, 
\begin{equation}
    s_{g+1/2} = \frac{\ln \left(\bar{Q}_{g+1}/\bar{Q}_{g}\right)}{\ln \left(\bar{\nu}_{g+1}/ \bar{\nu}_{g}\right)},
    \label{eq:sg}
\end{equation}
where $\bar{\nu}_g = \sqrt{\nuplus \numinus}$ is the bin centre in logarithmic space and $\bar{Q}_{g} = Q_{g} / (\nuplus - \numinus)$ is the average specific radiation quantity in a bin. Then, we define
\begin{equation}\label{eq:minmod}
    \alpha_{Q,g} = \begin{cases}
        -1, & g = 1 \ {\rm or} \ N_g \\
        {\rm minmod}(s_{g-1/2}, s_{g+1/2}) & g = 2, 3, \cdots, N_g - 1 \\
    \end{cases}
\end{equation}
Note here the special treatment of the two edge groups, $g = 1$ and $N_g$. We cannot treat these as we treat all other groups since we cannot compute edge slopes for them, and must instead pick specific values. 
We choose $\alpha_{Q,g} = -1$ for the edge groups because, as demonstrated in the main text, the quantity-weighted average of $\alpha_{Q,g}$ over all groups must be $-1$. We provide a visual demonstration of this method and compare it with the PPL-fixed reconstruction method in \autoref{fig:slopes}.

\bsp	%
\label{lastpage}
\end{document}